\documentclass[format=acmsmall,screen=true,review=false]{acmart}

\usepackage{caption}
\usepackage{subcaption}
\usepackage{graphicx}
\usepackage{amsmath}
\usepackage{amssymb}
\usepackage{textcomp}
\usepackage{booktabs}
\usepackage[justification=justified]{caption}
\makeatletter  
\newif\if@restonecol 
\makeatother

\makeatletter 
\newif\if@restonecol 
\makeatother

\usepackage[linesnumbered,ruled,vlined]{algorithm2e}
\usepackage{algpseudocode} 
\usepackage{amsmath} 
\usepackage{color} 
\usepackage{float}
\usepackage{hyperref} 
\usepackage{url}
\usepackage{multirow}

\AtBeginDocument{%
 \providecommand\BibTeX{{%
  \normalfont B\kern-0.5em{\scshape i\kern-0.25em b}\kern-0.8em\TeX}}}

\acmJournal{TOIT}
\setcopyright{acmcopyright}
\acmVolume{1}
\acmNumber{1} 
\acmArticle{1}
\acmYear{2022}
\acmMonth{3}

\begin{document}

\title[esDNN: Deep Neural Network based Multivariate Workload Prediction Approach in Cloud Environments]
{esDNN: Deep Neural Network based Multivariate Workload Prediction Approach  in Cloud Computing Environments}

\author{Minxian Xu}
\email{mx.xu@siat.ac.cn}
\author{Chenghao Song}
\email{ch.song@siat.ac.cn}
\affiliation{%
 \institution{Shenzhen Institute of Advanced Technology, CAS}
 \city{Shenzhen}
 \state{Guangdong}
 \country{China}
 \postcode{518000}
}

\author{Huaming Wu}

\affiliation{%
 \institution{Tianjin University}
 \city{Tianjin}
 \country{China}}
\email{whming@tju.edu.cn}

\author{Sukhpal Singh Gill}
\affiliation{%
 \institution{Queen Mary University of London}
 \city{London}
 \country{UK}
}

\author{Kejiang Ye}
\authornote{Corresponding author: Kejiang Ye}
\affiliation{%
\institution{Shenzhen Institute of Advanced Technology, CAS}
 \city{Shenzhen}
 \state{Guangdong}
 \country{China}}

\author{Chengzhong Xu}
\affiliation{%
 \institution{State Key Lab of IOTSC, University of Macau}
 \city{Macau}
 \country{China}}

\renewcommand{\shortauthors}{Xu et al.}

\begin{abstract}
 Cloud computing has been regarded as a successful paradigm for IT industry by providing benefits for both service providers and customers. In spite of the advantages, cloud computing also suffers from distinct challenges, and one of them is the inefficient resource provisioning for dynamic workloads. Accurate workload predictions for cloud computing can support efficient resource provisioning and avoid resource wastage. However, due to the high-dimensional and high-variable features of cloud workloads, it is difficult to predict the workloads effectively and accurately. The current dominant work for cloud workload prediction is based on regression approaches or recurrent neural networks, which fail to capture the long-term variance of workloads. To address the challenges and overcome the limitations of existing works, we proposed an \textit{\textbf{e}}fficient \textit{\textbf{s}}upervised learning-based \textit{\textbf{D}}eep \textit{\textbf{N}}eural \textit{\textbf{N}}etwork (\textbf{\textit{esDNN}}) approach for cloud workload prediction. Firstly, we utilize a sliding window to convert the multivariate data into supervised learning time series that allow deep learning for processing. Then we apply a revised Gated Recurrent Unit (GRU) to achieve accurate prediction. To show the effectiveness of esDNN, we also conduct comprehensive experiments based on realistic traces derived from Alibaba and Google cloud data centers. The experimental results demonstrate that esDNN can accurately and efficiently predict cloud workloads. Compared with the state-of-the-art baselines, esDNN can reduce the mean square errors significantly, e.g. 15\% than the approach using GRU only. We also apply esDNN for machines auto-scaling, which illustrates that esDNN can reduce the number of active hosts efficiently, thus the costs of service providers can be optimized.
\end{abstract}



\maketitle

	\section{Introduction} \label{sec:introdution}

	\color{black}Today's organizations and enterprises are becoming more dependent upon information technologies with cloud services that are deployed in cloud data centers \cite{XuBrownoutSurvey,Buyya2008}. Cloud services offer significant benefits for both customers and service providers \cite{ghahramani2017toward}. \color{black}The customers can access the services with high availability, and the service providers can take advantage of elasticity and low management costs of infrastructure. The pay-as-you-go pricing model is also a dominant benefit that promotes the fast development of cloud computing \cite{Du2020TC}. Due to these benefits, large cloud service providers, e.g. Amazon, Google and Microsoft have established large-scale data centers to provide resources for their services and a great number of companies have started to migrate their local services to the cloud \cite{Chen2019WWW}. 
	
	Although cloud computing is featured with these attractive benefits, some unpredictable situations, e.g. workload bursts can lead to resources being insufficient. The unmatched resources for workloads can also waste resources or degrade performance, for instance, more resources are provisioned than required when workloads are at a low level and only limited resources are offered when workloads are increasing dramatically \cite{Wang2020TC}. Therefore, to improve the resource usage, predicting workloads in an accurate manner is required. With the effective prediction of future workloads, the service provider can plan resources in a more efficient and rational way by allocating or de-allocating resources in advance \cite{ChenTPDS2020}. 
	
		However, it is not an easy job to predict cloud workloads efficiently and accurately due to their native characteristics. Cloud workloads have high variance and high dimensionality, which make them difficult to forecast. High variance represents that the number of workloads and their demanded resources can change dramatically. According to the analysis of Alibaba cloud data centers, the average
		resource utilization can range from 5\% to 80\% \cite{AlibabaTraces}. And in Google cloud data centers, workloads can change randomly during a specific observation period. As for high dimensionality, it represents that cloud workload traces record a great amount of information and different specification of machines, which needs to extract the necessary and valuable information for the training model.
		
		To address the high variance challenge of cloud workloads, the pattern of workloads, as well as the relationship with time series, should be learned and exploited to design efficient and accurate prediction algorithms to fit with the variances of workloads. As for the high dimensionality challenge, the dataset can be further analyzed to extract the necessary data while assuring the prediction accuracy.
		
		A significant amount of research has been devoted to cloud workload prediction. Traditional approaches are mostly based on the regression methods, heuristic algorithms and traditional neural network approaches. \textcolor{black}{Traditional neural networks generally refer to shallow networks that contain only several layers, such as Multi-layer Perceptron (MLP) and Radial Basis Function (RBF).} However, these approaches can only work effectively for the workloads with obvious patterns, e.g. for small-scale data centers for ordinary companies or organizations. For the large-scale public cloud data centers, these approaches can not obtain high prediction accuracy. The main reason is that the regression methods and simple neural networks cannot capture the complicated correlation of workloads. Therefore, to achieve higher accuracy, more complicated neural networks can be applied to take full advantage of the correlations of neurons. 
		
		As a representative of neural network-based approaches, the Recurrent Neural Network (RNN) \cite{Mikolov2011RNN} has been applied to predict cloud workloads as it has the feature to model the changes with time series. RNN can use its memory to process a set of inputs in sequence. However, it is inefficient for RNN to learn long-term memory dependencies because of the gradient vanishing. To overcome this limitation, some revised RNN, including Long Short-Term Memory (LSTM) \cite{LSTM} and Gated Recurrent Unit (GRU) \cite{GRU} have been proposed, which have demonstrated a strong capacity to learn long-term memory dependencies. 
Compared with LSTM, GRU has demonstrated better prediction accuracy and learning efficiency in practice. Thus, in this work, we apply a GRU-based approach to capture the variance of cloud workloads. 
	\subsection{Motivation and Our Contributions}
		\color{black}
To address the high dimensionality challenge, extraction of features of the original data is required. 
Our main motivations are as follows: 
\begin{itemize}
	\item Some approaches including Principal Component Analysis (PCA) \cite{PCA} and auto-encoder \cite{autoencoder} have been investigated, which can reduce dimension largely, while the accuracy is degraded as some features have been ignored. 
\item Traditional machine learning models can only show the mapping relationship between the source data and the target data, however, the time relationship cannot be extracted and exploited.
\item When predicting long periods, the dominant time-series data prediction approaches based on LSTM and RNN have the limitations of gradient disappearance and explosion. 
\end{itemize}

To address the aforementioned challenges for cloud workload prediction, we first extract some key features from the realistic traces derived from the cloud data center, and then convert the multivariate time series into supervised learning time series \cite{brownlee2016supervised} for further training with our designed training algorithm based on GRU. Our objective is to achieve efficient and accurate predictions for highly variable
and high dimensional cloud workloads to finally optimize the resource usage in cloud computing environments.

	The main \textbf{contributions} of this paper are summarized as follows:\color{black} 
	\begin{itemize}	
	
	\item The sliding window for Multivariate Time series Forecasting (S-MTF) is designed to convert multivariate time series into supervised learning time series for multivariate workloads and keep sufficient information. The S-MTF can reorganize the time series to sample X and label Y and model the correlation between predicted data, which can use algorithms based on Deep Neural Network (DNN) to achieve predictions.
	
	\item An \textit{\textbf{e}}fficient \textit{\textbf{s}}upervised learning-based \textit{\textbf{D}}eep \textit{\textbf{N}}eural \textit{\textbf{N}}etwork (\textbf{\textit{esDNN}}) algorithm is proposed for cloud workload prediction to learn and capture the features of historical data and accurately predict future workloads. The proposed algorithm can adapt to the variances of workloads by updating the gates of GRU and overcome the limitations of gradient disappearance and explosion. 

	\item Comprehensive experiments are conducted by using realistic data derived from Alibaba and Google cloud data centers to evaluate the performance of esDNN. The results demonstrate that the proposed approach can achieve better prediction accuracy than state-of-the-art algorithms. Experiments also show that the proposed approach can be applied for auto-scaling scenarios to improve resource provisioning. 

\end{itemize}
		\color{black}
\subsection{Article Organization}	
The rest of the paper is organized as follows: Section \ref{sec:related} discusses the related work for workload prediction in cloud computing environments. Section \ref{sec:problemmodel} depicts the system model of our proposed approach, followed by system problem statement. The proposed algorithm based on DNN is introduced in Section \ref{sec:algorithm}. Section \ref{sec:EXPERIMENT} introduces the details of our experiments that apply dataset derived from realistic traces to predict workloads, and demonstrate the feasibility of our approach to improve the resource provisioning of cloud data centers. Finally, conclusions along with the future directions are given in Section \ref{sec:conclusion}.
	
	\section{Related Work} \label{sec:related}
	
	Many researchers have conducted research on workload prediction. The main contributions for cloud workload prediction can be classified as regression-based and learning-based approaches. The regression-based approaches mainly include linear regression, auto-regression and other traditional regression-based approaches. While for the learning-based approaches, both the traditional approaches based on machine learning and some updated methodologies based on deep learning have been investigated.
	
	\subsection{Regression-based Approaches for Cloud Workload Prediction}
	Calheiros et al. \cite{Calheiros2015} proposed an approach based on auto-regression to predict future workloads by using requests of web applications. The proposed approach can achieve high accuracy in resource utilization and QoS prediction. Yang et al. \cite{Yang2014} introduced an approach based on linear regression for workload prediction to satisfy Service Level Agreement (SLA) and reduce scaling costs. Based on the prediction data, the auto-scaling mechanism can be further applied to optimize virtualized resource usage. Centinski et al. \cite{centinski2015} combined statistical and machine learning methods together to improve workload prediction for cloud applications. The training method is utilized to learn the dominant system parameters of the influence application, and the prediction method is based on the regression approach. Singh et al. \cite{Singh2019} presented a combined algorithm based on linear regression and support vector machine for workload prediction of web applications. A workload classifier was also proposed to select the model based on workloads features.  Liu et al. \cite{LIU2017JNCA} introduced an adaptive workload prediction approach based on workloads classification, in which different prediction models can be assigned to the different categorized workloads.  Bi et al. \cite{Bi2019} proposed a prediction method that integrates Savitzky-Golay filter and wavelet decomposition with stochastic configuration networks to predict workloads. 
	
	These regression-based approaches have proven their effectiveness in workload prediction. However, most of these approaches are only suitable for workloads with obvious patterns, e.g. Wikipedia workloads with fixed daily tendencies. The modern cloud workloads with high variance make these approaches hard to represent correlations between different parameters. Besides, these approaches were applied to high-performance computing workloads, small-scale data centers or synthetic workloads, which have lower variance compared with cloud workloads. Therefore, to efficiently capture the characteristics of cloud workloads, more advanced learning approaches, e.g. machine learning and deep learning-based methodologies have been investigated.

	\subsection{Learning-based Approaches for Cloud Workload Prediction}
	Kumar et al. \cite{KUMAR2018} applied a neural network and self-adaptive differential evolution algorithm to learn and extract the pattern from workloads. This evolution-based approach can reduce the prediction error by searching a large solution space, thereby minimizing the effects of initial solution choice. Zheng et al. \cite{ZhangTII2018} presented a deep learning model based on canonical polyadic decomposition to predict the usage of virtual machines for cloud workloads for industry informatics. Compared with machine learning-based approaches, deep learning-based approaches can achieve higher accuracy. Kumar et al. \cite{KumarICSCC2018} proposed a prediction model based on LSTM and showed good performance in reducing mean square errors. Qiu et al. \cite{Qiu2016} introduced a deep learning approach to predict Virtual Machine (VM) workloads by extracting high-level features of VMs workloads and then predicting future VM workloads. Zhu et al. \cite{zhu2019novel} presented an approach based on LSTM encoder-decoders network with an attention mechanism. The features of historical data are extracted via the encoder network and the attention mechanism is integrated into the decoder network. Amiri et al. \cite{Amiri2018} introduced an online learning approach to adapt resources according to workloads variations based on sequential pattern mining, which can learn new behavioral patterns rapidly. Chen et al. \cite{ChenTPDS2020} proposed a deep learning-based approach, which includes a top-sparse auto-encoder to extract essential features of workloads and GRU to obtain an accurate and adaptive prediction for cloud workloads. Several different types of workloads have been investigated to validate the effectiveness of the proposed approach. Eli et al. \cite{Eli2017} presented a resource central system to collect Azure VM parameters to learn the VM behavior offline with Microsoft learning libraries and then make online resource usage prediction, which predicts the oversubscription of VM types while ensuring VM performance.
	
	\color{black}
	Bi et al. \cite{bi2021integrated} applied bi-directional LSTM (Bi-LSTM) to predict large-scale workloads and resource consumption in the cloud computing environment. The performance of the approach has been validated with Google traces and shown better results than baselines. 
\color{black}		Karim et al. \cite{ karim2021bhyprec} proposed a hybrid approach combing RNN and Bi-LSTM to forecast CPU workload of VMs, which can improve the performance of using a single technique separately. 	
	Chen et al. \cite{ chen2021risk} introduced the LSTM-based approach to predict the useful life of components to indicate system health. A support vector regression is also combined to enhance the prediction robustness and marginal utility. Results based on NASA have validated the effectiveness of the proposed approach.
	\color{black}Singh et al. \cite{singh2021quantum} proposed an evolutionary quantum neural network-based approach for cloud workloads prediction, which leverages the computational efficiency of quantum computing to encode workloads, and utilizes the neural network to estimate resource demands. The experiments with traces from cloud data centers and traditional data centers have validated the effectiveness of the proposed approach. 	
	Kim et al. \cite{kim2020forecasting} introduced a cloud prediction framework named CloudInsight that combines multiple predictors based on traditional machine learning techniques to enable accurate predictions for real cloud workloads. The ensemble supports dynamic and periodical optimization to handle the variations of workloads. The framework can also reduce the periods of under-provisioning and over-provisioning, thus improving system efficiency. 
	\color{black}

	\color{black}The deep learning based approaches have been applied in predictions in many areas, such as communication, economic market, and pedestrian motion. Sun et al. \cite{sun2021lstm} proposed LSTM-based approach to predict link quality confidence interval for wireless communication under a smart grid environment. A wavelet denoising algorithm has been applied to decompose the signal-to-noise ratio time series into the deterministic and stochastic ones to train two LSTM neural networks. 
	Li et al. \cite{ li2020recurrent} introduced a recurrent attention and interaction model to predict pedestrian trajectories, which includes several modules to achieve precise prediction collaboratively. The introduced approach can comprehensively mine the spatio-temporal information to model attention mechanisms, interactions, and multimodality of pedestrian motion. 
	Barra et al. \cite{barra2020deep} presented an approach to forecast market behavior by encoding time series to Gramian angular fields images based on neural networks. Qiao et al. \cite{qiao2019self} proposed an approach based on a neural network to model the uncertain nonlinear systems by utilizing a distance concentration algorithm to increase prediction accuracy and reduce computation time. 
	However, these approaches are not focusing on cloud workloads prediction. 
	\color{black}
	
    To summarize, most of the learning-based approaches are based on machine learning algorithms or traditional RNN, which either cannot exploit the long-term memory dependencies or address the gradient vanishing challenge. Thus, it is also difficult for them to predict cloud workloads accurately. Only limited research has paid attention to GRU, which is an improved version of RNN and can address the gradient vanishing challenge to achieve better accuracy. For instance, Chen et al.~\cite{ChenTPDS2020} applied GRU for cloud workload prediction, however, they also apply the auto-encoder approach to compress the dimensionality of the original data. Although the auto-encoder approach can address the high dimensionality, the accuracy is also undermined since the full data is not utilized to capture the whole features of workloads.

		\subsection{Critical Analysis}
		\color{black}
	The current paper contributes to the growing body of work in the cloud workload prediction area. The comparison of our proposed approach and the related work is summarized in Table \ref{tab:relatedwork}. To solve the aforementioned challenges, {e.g. high-dimensional problems and multivariate problems,} we apply GRU to capture the long-term memory dependencies to address the high variance of cloud workloads, thereby achieving high accuracy prediction of cloud workloads. We also apply a sliding window for multivariate time series prediction to convert the original time series into supervised learning time series to address the high dimensionality and further achieve higher accuracy. From the technique perspective, our GRU-based approach is advanced in prediction compared with traditional regression and machine learning based approaches, and aims to overcome the limitation of gradient disappearance and explosion that exist in approaches like LSTM. From the data preprocessing perspective, our approach focusing on a sliding window to take advantage of full data information and the correlation between the predicted data rather than only extracting part of data like in auto-encoded based approach. We also validated our approach based on realistic traces of Google and Alibaba and multiple metrics have been evaluated comprehensively.
	
	\color{black}


\begin{table*}[]
\centering
\caption{Comparison of related work}
\label{tab:relatedwork}
\resizebox{\textwidth}{!}{%
\begin{tabular}{|c|c|c|c|c|c|c|c|c|c|c|c|c|c|c|c|c|c|c|c|}
\hline
\multirow{3}{*}{\textbf{Approach}} & \multicolumn{7}{c|}{\textbf{Technique}} & \multicolumn{2}{c|}{\textbf{Data Preprocessing}} & \multicolumn{3}{c|}{\textbf{Predicted Resources}} & \multicolumn{3}{c|}{\textbf{Workloads}} & \multicolumn{4}{c|}{\textbf{Performance Metrics}} \\ \cline{2-20} 
 & \multirow{2}{*}{\textbf{Regression}} & \multirow{2}{*}{\textbf{\begin{tabular}[c]{@{}c@{}}Machine \\ Learning\end{tabular}}} & \multirow{2}{*}{\textbf{\begin{tabular}[c]{@{}c@{}}Episode\\ Mining\end{tabular}}} & \multicolumn{4}{c|}{\textbf{\begin{tabular}[c]{@{}c@{}}Deep \\ Learning\end{tabular}}} & \multirow{2}{*}{\textbf{Auto-encoder}} & \multirow{2}{*}{\textbf{\begin{tabular}[c]{@{}c@{}}Sliding \\ window\end{tabular}}} & \multirow{2}{*}{\textbf{\begin{tabular}[c]{@{}c@{}}VM \\ utilization\end{tabular}}} & \multirow{2}{*}{\textbf{\begin{tabular}[c]{@{}c@{}}Server \\ Utilization\end{tabular}}} & \multirow{2}{*}{\textbf{QoS}} & \multicolumn{2}{c|}{\textbf{Realistic}} & \multirow{2}{*}{\textbf{Synthetic}} & \multirow{2}{*}{\textbf{MSE}} & \multirow{2}{*}{\textbf{RMSE}} & \multirow{2}{*}{\textbf{MAPE}} & \multirow{2}{*}{\textbf{CDF}} \\ \cline{5-8} \cline{14-15}
 &  &  &  & DNN & DBN & LSTM & GRU &  &  &  &  &  & \begin{tabular}[c]{@{}c@{}}Cloud \\ Data Centers\end{tabular} & \begin{tabular}[c]{@{}c@{}}Traditional \\ Data Centers\end{tabular} &  &  &  &  &  \\ \hline
Calheiros et al. \cite{Calheiros2015} & $\surd$ &  &  &  &  &  &  &  &  &  & $\surd$ & $\surd$ &  & $\surd$ &  &  &  & $\surd$ &  \\ \hline
Amiri et al. \cite{Amiri2018} &  &  & $\surd$ &  &  &  &  &  &  & $\surd$ &  & $\surd$ &  & $\surd$ & $\surd$ &  &  & $\surd$ &  \\ \hline
Ceninski et al. \cite{centinski2015} &  & $\surd$ &  &  &  &  &  &  &  &  &  & $\surd$ &  & $\surd$ &  & $\surd$ &  &  &  \\ \hline
Kumar et al. \cite{KUMAR2018} &  &  &  & $\surd$ &  &  &  &  &  & $\surd$ &  & $\surd$ &  & $\surd$ &  & $\surd$ &  &  &  \\ \hline
Kumar et al. \cite{KumarICSCC2018} &  &  &  &  &  & $\surd$ &  &  &  &  & $\surd$ &  &  & $\surd$ &  & $\surd$ &  &  &  \\ \hline
Qiu et al. \cite{Qiu2016} &  &  &  &  & $\surd$ &  &  &  &  & $\surd$ & $\surd$ &  &  & $\surd$ & $\surd$ &  &  & $\surd$ &  \\ \hline
Rodrigo et al. \cite{Calheiros2015} & $\surd$ &  &  &  &  &  &  &  &  & $\surd$ &  & $\surd$ &  & $\surd$ &  &  &  & $\surd$ & $\surd$ \\ \hline
Singh et al. \cite{Singh2019} & $\surd$ & $\surd$ &  &  &  &  &  &  & $\surd$ & $\surd$ &  & $\surd$ &  & $\surd$ &  & $\surd$ & $\surd$ & $\surd$ &  \\ \hline
Yang et al. \cite{Yang2014} & $\surd$ &  &  &  &  &  &  &  &  & $\surd$ & $\surd$ &  &  & $\surd$ & $\surd$ &  &  &  & $\surd$ \\ \hline
Zhang et al. \cite{ZhangTII2018} &  & $\surd$ &  &  &  &  &  & $\surd$ &  & $\surd$ &  &  &  & $\surd$ &  &  & $\surd$ & $\surd$ &  \\ \hline
Zhu et al. \cite{zhu2019novel} &  &  &  &  &  & $\surd$ &  & $\surd$ &  &  & $\surd$ &  & $\surd$ &  &  & $\surd$ & $\surd$ & $\surd$ &  \\ \hline
Bi et al. \cite{Bi2019} &  &  &  & $\surd$ &  &  &  &  &  &  & $\surd$ &  & $\surd$ &  &  & $\surd$ &  &  &  \\ \hline
Liu et al. \cite{LIU2017JNCA} & $\surd$ &  &  &  &  &  &  &  &  &  & $\surd$ &  & $\surd$ &  &  & $\surd$ &  &  &  \\ \hline
Eli et al. \cite{Eli2017} &  & $\surd$ &  &  &  &  &  &  &  & $\surd$ & $\surd$ &  & $\surd$ &  &  &  &  &  & $\surd$ \\ \hline
\color{black}
Sun et al. \cite{sun2021lstm} & & & & & & \color{black}$\surd$ &  & & & & & \color{black}$\surd$ & & &\color{black} $\surd$ & \color{black} $\surd$ & & & \\ \hline
\color{black}
Li et al. \cite{li2020recurrent}  & & & & & & \color{black} $\surd$ &  & & & & & & & & &  &  & & \\ \hline

\color{black}Barra et al. \cite{barra2020deep} & & & & \color{black}$\surd$ & & & &  & & & & & & & & & & & \\ \hline

\color{black}Qiao et al. \cite{qiao2019self} & & & & \color{black}$\surd$ & & & & & & & & & & & & & \color{black}$\surd$ & & \\ \hline

\color{black}Bi et al. \cite{bi2021integrated} & \color{black}$\surd$ & & & & & \color{black}$\surd$ & & & & & \color{black}$\surd$ & \color{black}$\surd$& \color{black}$\surd$ & & & \color{black}$\surd$ & & & \\ \hline

\color{black}Singh et al. \cite{singh2021quantum} & & & & \color{black}$\surd$& & & & & & & \color{black}$\surd$ & & \color{black}$\surd$ & \color{black}$\surd$ & & & \color{black}$\surd$& & \\ \hline

\color{black}Kim et al. \cite{kim2020forecasting} & & \color{black}$\surd$ & & & & & & & & & \color{black}$\surd$ & \color{black}$\surd$& \color{black}$\surd$ & \color{black}$\surd$ & & & \color{black}$\surd$ & \color{black}$\surd$ & \color{black}$\surd$ \\ \hline

\color{black}Karim et al. \cite{ karim2021bhyprec}  & &  & & & & \color{black}$\surd$ & \color{black}$\surd$& & \color{black}$\surd$ & \color{black}$\surd$ &  & &  & \color{black}$\surd$ & & \color{black}$\surd$ & \color{black}$\surd$ & \color{black}$\surd$ &  \\ \hline

\color{black}Chen et al. \cite{ chen2021risk}  & \color{black}$\surd$ &  & & & &  \color{black}$\surd$ & & &  &  &  &  &  &  \color{black}$\surd$ & &  & \color{black}$\surd$ &  &  \\ \hline

esDNN (This Work) &  &  &  &  &  &  & $\surd$ &  & $\surd$ &  & $\surd$ &  & $\surd$ &  &  & $\surd$ & \color{black}$\surd$ & \color{black}$\surd$ & $\surd$ \\ \hline
\end{tabular}%
}
\end{table*}

\section{System Model} \label{sec:problemmodel}

In this section, we introduce our system model and optimization objective. In our system model, we aim to offer an efficient and accurate prediction model that the service providers can apply to predict future workloads. Thus, the resource usage can be optimized to reduce their costs, e.g. integrating the model with auto-scaling to reduce the number of active hosts.  

	\begin{figure*}[ht]
		\centering
		\includegraphics[width=0.5\linewidth]{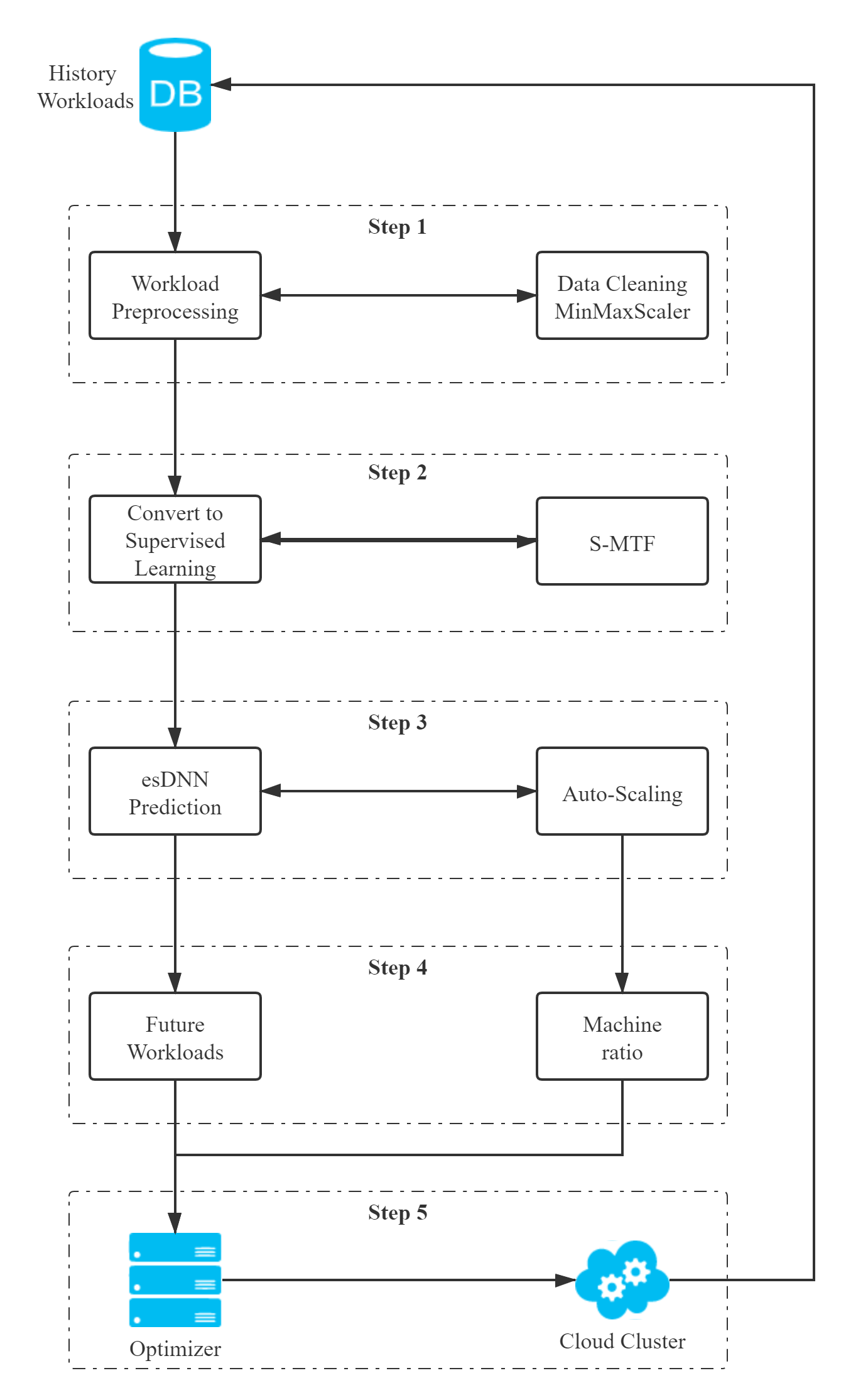}
		\caption[VarPerOptCom]{Multivariate time series prediction model for cloud  workloads}
		\label{fig:Model}
	\end{figure*}
	
It is not easy to predict cloud workloads as they can change dramatically within a short time and the pattern is also difficult to capture precisely. \textcolor{black}{For example, workloads in every 5 minutes from the dataset of Alibaba can vary significantly \cite{AlibabaTraces}.} The cloud workloads are tightly coupled with time series, and it is inefficient to get accurate prediction results from a simple regression model or univariate time predictions. Multivariate time series can contain more dynamic information than univariate time series. For instance, the data in multivariate time series forecasting can have certain correlations, such as CPU usage and memory usage in workload forecasting. Therefore, we built a multivariate time series forecasting model to predict highly random workloads and use the real-world dataset to verify the accuracy of the model. In our prediction model, we use CPU usage as our standard for measuring the prediction results. Fig.~\ref{fig:Model} shows the main components and flow of the system model.
	
	\textbf{Step 1: Data Preprocessing.} This step is equipped with workloads preprocessing component and data cleaning component, which processes the raw data derived from the realistic cloud traces. With the raw data of cloud workloads, we first remove the columns that contain empty data. Because whether it is to use the zero-filling scheme or simply ignore these data, they will have a negative impact on our forecast data. Afterwards, we classify the dataset by time, then calculate the average value of each parameter with the same timestamp. Next, we normalized the Alibaba dataset and Google dataset. Normalization is a dimensionless processing method that makes the absolute value of the physical system value into a certain relative value relationship. From the perspective of model optimization, normalization can not only improve the convergence speed of the model but also improve the accuracy of prediction. The normalization method has two forms, one is to change the number to a decimal between (0, 1), and the other is to change the dimensional expression to a non-dimensional expression and become a scalar. In this article, the former is chosen as the normalization method, and we use MinMaxScaler to achieve this function. The MinMaxScaler operation is based on the min-max scaling method as follows: 
	\begin{align}
X_\text{std}&=\frac{X-X_{\min }}{X_{\max }-X_{\min }}-X_{\min },\\
X_\text{scaled}&=X_\text{std} *(max - min) + min.
\end{align}
	\textcolor{black}{We apply the MinMaxScaler to transform features with default configurations and scale each feature to be a value between $min$ and $max$. The $X$ represents the set of the data to be processed, and the $X_{\min}$ and $X_{\max}$ are the minimum and maximum data in the dataset, and the final processed data is represented by $X_\text{scaled}$. To be noted, in this work, the predicted value is resource utilization, which ranges from  0.0 to 1.0, and the MinMaxScaler can handle these data well. As for the missing data, they will be filled with the data in the previous time slot.}

\textbf{Step 2: Supervised Learning Conversion.} The difference between supervised learning and unsupervised learning lies in whether there are labels of samples for training. In supervised learning, it has labeled training samples. It trains through the existing training samples to obtain an optimized model and then uses this model to map all inputs to the corresponding outputs, thereby realizing data prediction and classification. For unsupervised learning, there are no pre-labeled training samples. 
In our system model, we use the supervised learning transfer function to convert the multivariate time series prediction problem into a supervised learning problem based on \cite{brownlee2016supervised}. More details about the transfer function will be introduced in Section \ref{sec:algorithm}. The key motivation is to use the normalized dataset as the input of the transfer function, and reframe the time series datasets as supervised learning datasets. To achieve this, we split the dataset into a training set and a validation set. After that, the dataset is divided into sample X and its corresponding label Y. With these conversion operations, we can transform the time series forecasting problem into a supervised learning-based time series problem.

	\textbf{Step 3: Model Construction.} In this step, our system model focuses on the construction of deep learning networks and establishes an optimization model for cloud workload prediction based on the preprocessed data. The preprocessed data are considered as input and the output is the optimized parameters of the model as well as the evaluation metrics, e.g. mean square errors. In this step, the hyperparameters of the deep learning network should also be defined, e.g. the number of layers, number of neurons, and types of network. Our proposed network model is derived from GRU and more design details will be given in the following sections. 
By predicting the cloud workloads, we aim to obtain future resource usage and thus the optimization of the number of active machines can be optimized by auto-scaling approaches, which will be coordinately achieved with the next step.

\textbf{Step 4 and Step 5: Model Deployment and System Adaption.} These two steps focus on utilizing the models for workload prediction or other system optimization purposes. In the actual workload prediction, there is a time interval between two consecutive predictions, which means that the sequence prediction is based on a discrete time series dataset. For the Alibaba dataset, the prediction interval is usually 10 seconds, while the  time prediction interval of Google is 5 minutes. 
We apply this time interval as the prediction unit. Based on the trained model in Step 3, in this step, the system model can obtain the predicted future workloads and adjust the number of machines by applying auto-scaling. With the predicted data, the realistic system can dynamically adapt the resource provisioning for the system, e.g. adding or removing machines physically, which requires the use  of Application Programming Interfaces (APIs) provided by hardware.

\section{esDNN: Efficient Supervise Learning-based Deep Neural Network} \label{sec:algorithm}
This section presents our proposed approach, which is a deep learning-based approach for cloud workload prediction. To achieve efficient and accurate prediction results, a sliding window-based approach for multivariate time series prediction is applied to convert the original dataset into supervised learning-based time series data. Thereafter, a GRU-based deep learning network, named esDNN is proposed for future workload prediction.

        \subsection{Sliding Window for Multivariate Time Series Forecasting}
        
		\color{black}Time series forecasting requires the dataset to contain a set of time-dependent data, regardless of whether the time units of the dataset are seconds, minutes, or hours. This data needs to have a minimum time unit, but it does not need to have the same time interval between two adjacent timestamps. Having clarified this concept, we can say that a time series is a sequence of numbers sorted by time index. However, only have one time series is not sufficient. 	 		
		In Section \ref{sec:problemmodel}, we have introduced the definition of supervised learning. Complete supervised learning requires a sample group (X) and a label group (Y). There are two major differences compared with the work based sliding window \cite{dietterich2002machine} including 1) we consider the multivariate workloads to construct time series data rather than single variate; 2) we utilize the relationship between the predicted data by merging the predicted data and source data into supervised time series data together.
		To illustrate the conversion process more vividly, we use a small piece of sample data in the Alibaba cloud workloads dataset to show the conversion process and results. To simplify the example, we use one-step univariate forecasting. \color{black} 

			\begin{table}[]
			\centering
			\caption{One-step univariate forecasting: raw dataset}
			\label{tab:One-step univariate forecasting: Raw dataset}
            \begin{tabular}{|c|c|}
            \hline
            \textbf{Time} & \textbf{CPU utilization percentage} \\ \hline
            0              & 16.127                       \\ \hline
            10            & 21.5878                     \\ \hline
            20            & 17.3193                     \\ \hline
            30            & 16.8287                     \\ \hline
            40            & 18.6518                     \\ \hline
            \end{tabular}
            \end{table}
			 Table \ref{tab:One-step univariate forecasting: Raw dataset} shows the first five rows of data from the Alibaba dataset, after we convert these time series data into supervised learning data, it will be presented in the form of Table \ref{tab:One-step univariate forecasting: Supervised learning sequence}, where each row data is moved up with data in the group (Y)  and the time label has been removed.

			\begin{table}[]
				\centering
			\caption{One-step univariate forecasting: supervised learning sequence}
			\label{tab:One-step univariate forecasting: Supervised learning sequence}
				\begin{tabular}{|c|c|}
				\hline
				\textbf{X} & \textbf{Y} \\\hline 
				None          & 16.127                      \\\hline
				16.127        & 21.5878                     \\\hline
				21.5878       & 17.3193                     \\\hline
				17.3193       & 16.8287                     \\\hline
				16.8287       & 18.6518                     \\\hline
				18.6518       & None      \\\hline                  
				\end{tabular}
			\end{table}
For the multivariate time series datasets, we can also convert them into supervised learning datasets with sliding window approach. Similarly, we also take a small fragment from the Alibaba dataset. The difference is that in addition to Time and CPU utilization percent, we also take memory utilization percentage to reflect that this is a multivariate dataset. Here, we choose the memory utilization percentage as Y, which is considered as the label. In our model, we choose to use the one-step multivariate forecasting. Table \ref{Raw} and Table \ref{Supervised} show the original data and the converted data, respectively.

\begin{table}[]
	\centering
\caption{One-step multivariate forecasting: raw dataset}
\label{Raw}
\begin{tabular}{|c|c|c|}
\hline
\textbf{Time} & \textbf{CPU util. percentage} & \textbf{Memory util. percentage} \\ \hline
0    & 16.127             & 87.139            \\ \hline
10            & 21.5878                     & 87.0543                     \\ \hline
20            & 17.3193                     & 86.9491                     \\ \hline
30            & 16.8287                     & 86.9454                     \\ \hline
40            & 18.6518                     & 86.9495                     \\ \hline
50            & 20.0232                     & 86.9985                     \\ \hline
60            & 17.8671                     & 86.9249                     \\ \hline
\end{tabular}
\end{table}

\begin{table}[]
	\centering
\caption{One-step multivariate forecasting: supervised learning sequence}
\label{Supervised}
\begin{tabular}{|c|c|c|c|}
\hline
X1      & X2      & X3      & Y       \\\hline
None    & None    & 16.127  & 87.139  \\\hline
16.127  & 87.139  & 21.5878 & 87.0543 \\\hline
21.5878 & 87.0543 & 17.3193 & 86.9491 \\\hline
17.3193 & 86.9491 & 16.8287 & 86.9454 \\\hline
16.8287 & 86.9454 & 18.6518 & 86.9454 \\\hline
18.6518 & 86.9495 & 20.0232 & 86.9985 \\\hline
20.0232 & 86.9985 & 17.8671 & 86.9249 \\\hline
17.8671 & 86.9249 & None    & None   \\\hline
\end{tabular}
\end{table}

\begin{figure*}[ht]
		\centering
		\includegraphics[width=0.75\linewidth]{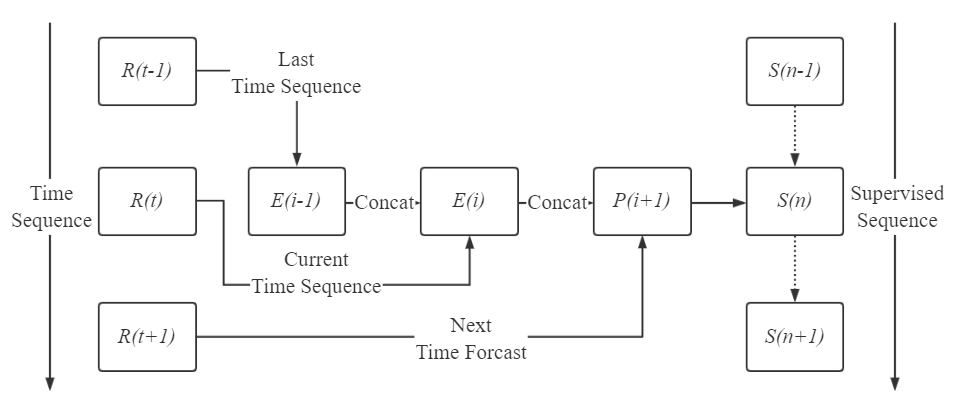}
		\caption[VarPerOptCom]{\color{black}Data conversion in S-MTF algorithm}
		\label{fig:S-MTF layer}
	\end{figure*}

Fig.~\ref{fig:S-MTF layer} shows the typical conversion process by operating the original data. Assuming that we have the time sequence as $R(t-1), R(t)$ and $R(t+1)$, where $R(t-1)$ is the last one, $R(t)$ is the current one and $R(t+1)$ is the next one. We set elements $E(i-1), E(i)$ and $P(i+1)$ as the elements to be combined as the supervised time sequence $S(n)$. The $E(i-1)$ is from the data of $R(t-1)$, $E(i)$ is assigned by $R(t)$, and $P(i+1)$ is assigned by $R(t+1)$. The $E(i-1)$ and $E(i)$ will be the sample data and $P(i+1)$ will be the label. The other supervised time sequence, e.g. $S(n-1)$ and $S(n+1)$ can be obtained in the same way.

\begin{figure*}[ht]
		\centering
		\includegraphics[width=0.7\linewidth]{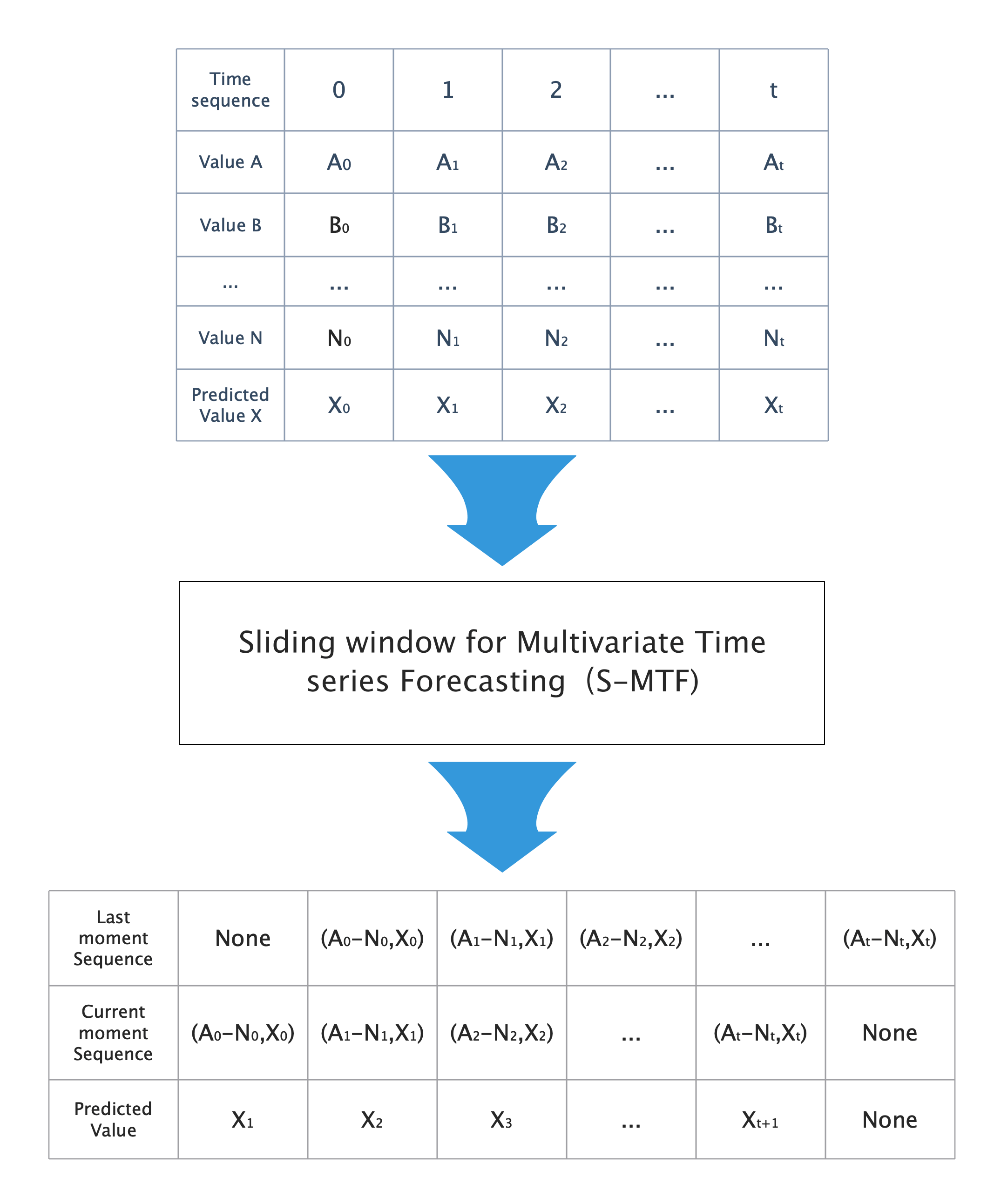}
		\caption[VarPerOptCom]{The key procedures of the S-MTF algorithm}
		\label{fig:SMTF}
		\end{figure*}

By this step, we have completed the application expression of multivariate time series forecasting, and what we have to do now is to abstract it into an algorithm. First, we define this algorithm as Sliding window for Multivariate Time series Forecasting (S-MTF) algorithm, which transforms a multivariate time series forecast into a supervised learning time series. The S-MTF algorithm can be applied to any time-related dataset, and it is still linearly related to time because it contains all the data of the previous moment at any time. At this point, the S-MTF is somewhat similar to LSTM, but the difference is that the forget gate of LSTM will weaken the influence from the previous moment, while S-MTF retains all the values of the previous moment, and it can be determined if you need to keep it or not. Besides, S-MTF contains the future label while using the multi-step forecasting. Furthermore, S-MTF satisfies the definition of supervised learning as it transforms time-related datasets into the sample and labeled datasets. With a more general form, Fig.~\ref{fig:SMTF} depicts the transformation of the S-MTF algorithm for time series and presents the supervised learning sequences obtained from the transformation in a tabular form.
Algorithm \ref{S-MTF} shows the pseudocode of the S-MTF algorithm. Before the original data are processed as the input of the algorithm, we have deleted NONE values in the dataset for the convenience of data processing, as they can influence the accuracy of the proposed algorithm. 

\color{black}
\textbf{Algorithm complexity analysis:} Given that there is a set of time series data with size $N$, the algorithm processes the data from 1 to $N-1$ to construct the matrix $S_n$. To obtain all the data in $S_n$ with 3 sub-data in each time interval, the complexity will be $O(3\times(N-1))$, which equals $O(N)$.  
\color{black}
    
	\begin{algorithm}[t]

		\color{black}
		\footnotesize
		\caption{Sliding window for Multivariate Time series Forecasting (S-MTF)} 
		\label{S-MTF}
		\SetAlgoLined
		\SetKwInOut{Input}{Input}\SetKwInOut{Output}{Output}
		\SetKwFor{ForAll}{forall }{do}{end forall}	
		
		\Input{Multivariate time series dataset $R_n$, $R_n$ contains a time series $R(1), R(2), \ldots, R(t), R(t+1)$, $k$ time-related variables, and a dataset $P_n$ to be predicted}
		\Output{Supervised learning dataset $S_n$, each row of it has $2k+3$ data} 
		Initialize an empty matrix $S_n$ to record supervised time series data
		\For{$t$ from $1$ to $n-1$}{

		     $E(t)$ $\leftarrow$ $R(t)$

		     \uIf{$t = 1$}{
		     Record NONE}
		     \Else{
		     $E(t-1)$ $\leftarrow$ $R(t-1)$
		    }
		    
		    \uIf{$t = n-1$}{
		     Record NONE}
		     \Else{
		     $P(t+1)$ $\leftarrow$ $R(t+1)$
		    }
		    
		    \textcolor{black}{Put $E(t-1)$, $E(t)$ and $P(t+1)$ together into a tuple} 
		    
		    Set $E(t-1)$, $E(t)$ as data in supervised  time series
		    
		    Set $P(t+1)$ as label in supervised  time series

		    $S(t)$ $\leftarrow$ \{$E(t-1)$, $E(t)$, $P(t+1)$\}
		    
		    \uIf{$S(t)$ contains NONE}
		    {Delete $S(t)$}
		    
		    Add $S(t)$ into $S_n$
        }
		    
	\end{algorithm}

    \subsection{esDNN Algorithm}
      \color{black}
      In our deep learning networks, the input data in the training phase include the resource utilization and the corresponding time series data, e.g. at time 08:00:00 am, the CPU utilization is 20\%. In the prediction phase, the input data are the resource utilization in the recent time intervals, e.g. the previous 5 minutes (can be configured via parameters in network model).
      \color{black}
      To construct our network model, we include one layer of Convolutional Neural Network (CNN). The CNN model is usually built on the feedforward neural network model. It generally consists of Input Layer, Convolutional Layer, Pooling Layer, Non-linearity Layer and Fully Connected Layer \cite{krizhevsky2012imagenet}. Two-dimensional convolutional neural networks (2D CNN) are widely used in image recognition, and one-dimensional convolutional neural networks (1D CNN) are generally used in Natural Language Processing (NLP). Additionally, the one-dimensional convolutional neural networks also have the capability in processing continuous sequences. For example, when obtaining a certain feature from a shorter segment in the whole dataset, while the feature is not highly correlated with the position of the data segment in the overall dataset, in this situation, the 1D CNN can play an important role. \color{black}The 1D CNN can extract features from local original time series data, and then model the short-term correlation between local time series data and subsequent trends \cite{xu2020tensorized}. \color{black}So we will use the 1D CNN to analyze our data. We built a one-dimensional convolutional layer and added it to our neural network. We also add padding, which maintains the boundary information of the time series. If there is no padding, most of the obtained information will only be operated by the convolution kernel once, but the data in the middle of the sequence are scanned many times, thus the results obtained will lose the accuracy of the boundary information. To improve accuracy, we apply a casual strategy for padding, which simply pads the layer's input with zeros in the front so that we can also predict the values of early time steps in the frame~\cite{sklearn_api}. Finally, we adopt Rectified Linear Unit (ReLU) as the activation function of the 1D CNN. After introducing the convolutional layer, the GRU-based layer can be added.

       GRU is a derived version of RNN. RNN uses traditional backpropagation and gradient descent algorithms to learn the target data. The BackPropagation Through Time (BPTT) algorithm is a commonly used method of training RNN. The idea of BPTT is the same as the backpropagation algorithm, which continuously finds better points along the negative gradient direction of the parameters to be optimized until convergence. However, the application of the BPTT algorithm can lead to the accumulation of activation function derivatives, which in turn leads to the occurrence of gradient disappearance and gradient explosion. In order to solve this problem, we can use two methods to avoid gradient explosion/disappearance. The first method is to replace the activation function. In our model, we avoid the disappearance of the gradient to a certain extent by setting ReLU as the activation function. But the derivative of ReLU in the range greater than 0 is always 1, which is easy to cause gradient explosion. Therefore, the second method is applied to change the circulation structure. 
    GRU that combines the forget gate and input gate into a single ``update gate'' is exploited in our model. It also merges cell state and hidden state and makes some other changes. 

    \begin{figure*}[ht]
		\centering
		\includegraphics[width=0.7\linewidth]{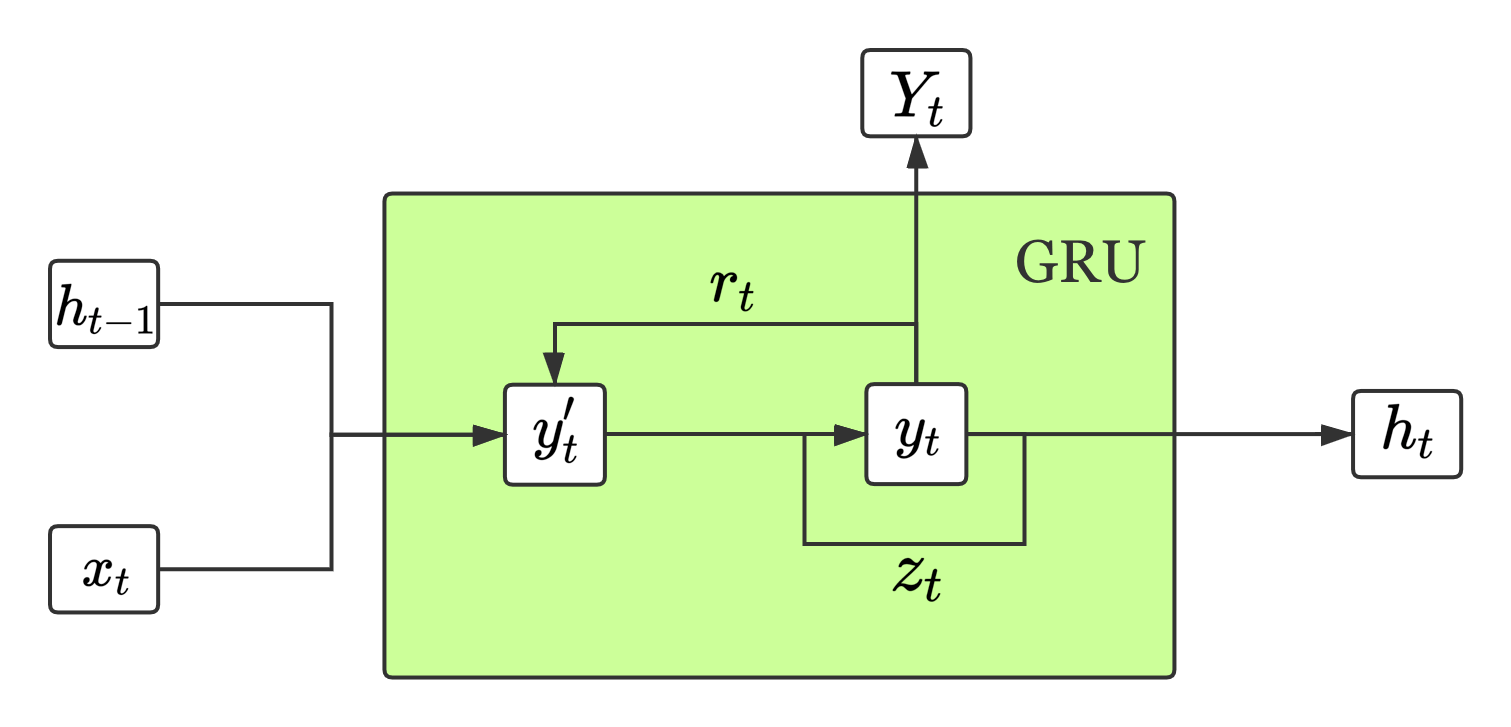}
		\caption[VarPerOptCom]{\color{black}GRU structure}
		\label{fig:GRU structure}
	\end{figure*}



    We have adopted the GRU-based neural network and made some improvements to optimize its performance in long sequence prediction. The structure of GRU is demonstrated in Fig.~\ref{fig:GRU structure}. The reset gate $r_t$ and update gate $z_t$ are the same as LSTM. But there is no output gate in GRU. Compared with LSTM, there is one less "gating" inside the GRU, which has fewer parameters than LSTM, but it can also achieve functions equivalent to LSTM \cite{fu2016using}. 

The sparse processing provided by ReLU can reduce the effective capacity of the model, which means too much feature masking makes the model unable to learn effective features. Since the gradient of ReLU is 0 when $x <0$, this neuron may never be activated by any data, which is called neuron necrosis. In addition, one of the similarities between ReLU and Sigmoid is that the result is a positive value without a negative value. To address this issue, we multiply ReLU and Sigmoid, and we can get the activation function Swish that is represented as below:
    \begin{equation}
        f(x)=x \cdot \operatorname{sigmoid}(\beta x),
    \end{equation}
    where $\beta$ is either a constant or a trainable parameter \cite{ramachandran2017searching}. 
    
    In the choice of activation function, instead of choosing the ReLU activation function that is commonly used by DNNs, we use Swish, which is a smooth and non-monotonic function. Its design is inspired by the use of sigmoid function for gating in LSTM. We use the same value for gating to simplify the gating mechanism, which is called self-gating. The advantage of self-gating is that it only requires a simple scalar input, while traditional gating requires multiple scalar inputs. This feature allows Swish to easily replace activation functions that take a single scalar as input without changing the hidden capacity or number of parameters. The pseudocode of esDNN is shown in Algorithm \ref{A:esDNN}.
    
    \color{black}
    \textbf{Algorithm complexity analysis:} The time complexity of esDNN depends on the number of networks ($N$), number of network weight connections ($C$), number of input node ($n$), hidden nodes ($h$), where $h \approx n$, dropout value ($d$). Therefore, the total time complexity for a maximum number of $b$ iterations is represented as $b \times O(n^2\times N \times C \times d)$, which equals to $O(n^2bdNC)$.   
    \color{black}
    
        \begin{algorithm}[t]
		\color{black}
		\footnotesize
		\caption{efficient supervise learning-based Deep Neural Network (esDNN)}
		\label{A:esDNN}
		\SetAlgoLined
		\SetKwInOut{Input}{Input}\SetKwInOut{Output}{Output}
		\SetKwFor{ForAll}{forall }{do}{end forall}	
		
		\Input{Multivariate time series dataset $R_n$.}
	\textbf{Hyperparameter}:time sequence $t$, training epochs $b$
		
	After processing by the S-MTF algorithm, multivariate time series dataset $R_n$ is transformed into the supervised learning dataset $S_n$
	
	Divide $S_n$ into training set $Ts$ and validation set $Vs$
    
    \textbf{esDNN:}
    Conv1D(filters=32, kernel\_size=5)
    
    \For{$i$ range from $0$ to $t$}{
    GRU:\\
    Update the reset gate $r_t$: \\
    $r_{t}=\sigma\left(W_{r} \cdot\left[h_{t-1}, x_{t}\right]\right)$ \\
    Update the update gate $z_t$: \\
    $z_{t}=\sigma\left(W_{z} \cdot\left[h_{t-1}, x_{t}\right]\right)$ \\
    Calculate the candidate hidden layer ${y}^\prime_t$: \\
    ${y}^\prime_{t}=\tanh \left(W \cdot\left[r_{t} * h_{t-1}, x_{t}\right]\right)$ \\
    Compute the output gate $y_t$:\\
    $y_{t}=\left(1-z_{t}\right) * h_{t-1}+z_{t} * \tilde{h}_{t}$
    }
    Dense(16, activation=``swish'')
    
    Dropout(0.2)
    
    Dense(1)
    
    \For{$j$ range from $0$ to $b$}{
    Train $Ts$ with esDNN, compare with $Vs$ 
    }
	  
	\end{algorithm}

\section{Performance Evaluation} \label{sec:EXPERIMENT}
In this section, we will firstly introduce the details about the dataset we use and the experimental configurations for workload prediction. Then we compare the performance of esDNN and other RNN-based approaches. Finally, we demonstrate that our approach can be applied to auto-scaling for cloud resource provisioning optimization.

	\subsection{Datasets and Environment Configuration}
We implement the multivariate time series forecasting based on TensorFlow 2.2.0 \cite{Tensorflow}, and the Python version is 3.7. We used two real-world datasets in the experiments for performance evaluation of our proposed approach. 
\begin{itemize}
\item \textbf{Alibaba dataset} \cite{AlibabaTraces}: It is cluster-trace-v2018 of Alibaba that recording the traces in 2018. Cluster-trace-v2018 includes about 4,000 machines in a period of 8 days, which we use all the data to make predictions. \textcolor{black}{The data can be found from Github}\footnote{https://github.com/alibaba/clusterdata}.
\item  \textbf{Google dataset} \cite{clusterdata:Wilkes2011}: It is derived from Google's cluster data-2011-2 recorded in 2011. The cluster data-2011-2 trace includes 29 days of data that contains 37,747 machines, including three different machine types. \textcolor{black}{The data can also be fetched from Github}\footnote{https://github.com/google/cluster-data}.
\end{itemize}

Both of these datasets can represent random features of cloud workloads. We use CPU usage as a key performance measurement of the accuracy of our prediction model. To show the effectiveness of prediction and remove the redundancy information, we configure the prediction time interval as 5 minutes. For the metadata for prediction, there are some differences between the two datasets because of the different types of data collected. For the Alibaba dataset, in addition to the time series and CPU usage data, we also select memory usage, incoming network traffic, outgoing network traffic, and disk I/O usage as the source data for prediction. As for Google dataset, in addition to the time series and CPU usage data, we also select canonical memory usage, assigned memory usage, total page cache memory usage as the source data for prediction. When processing the dataset from Google, we select 5 minutes as the time interval. Then we group the tasks according to the Machine ID and finally normalize the dataset.

\begin{figure*}[t]
	\centering
	\begin{subfigure}[b]{0.49\linewidth}
		\centering
		\includegraphics[height=4cm, width=0.99\linewidth]{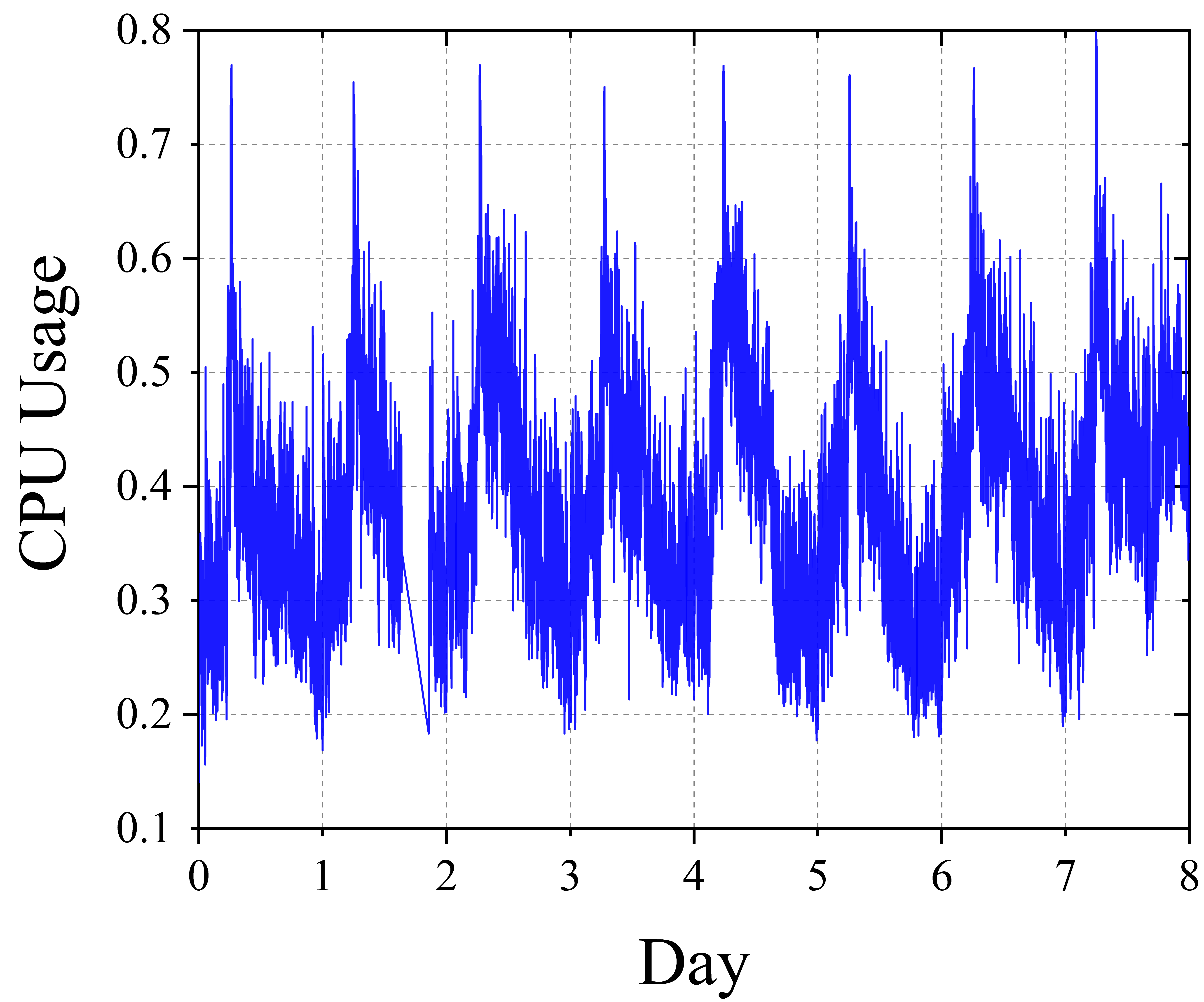}
	\end{subfigure}
	\begin{subfigure}[b]{0.49\linewidth}
		\centering
		\includegraphics[height=4cm, width=0.99\linewidth]{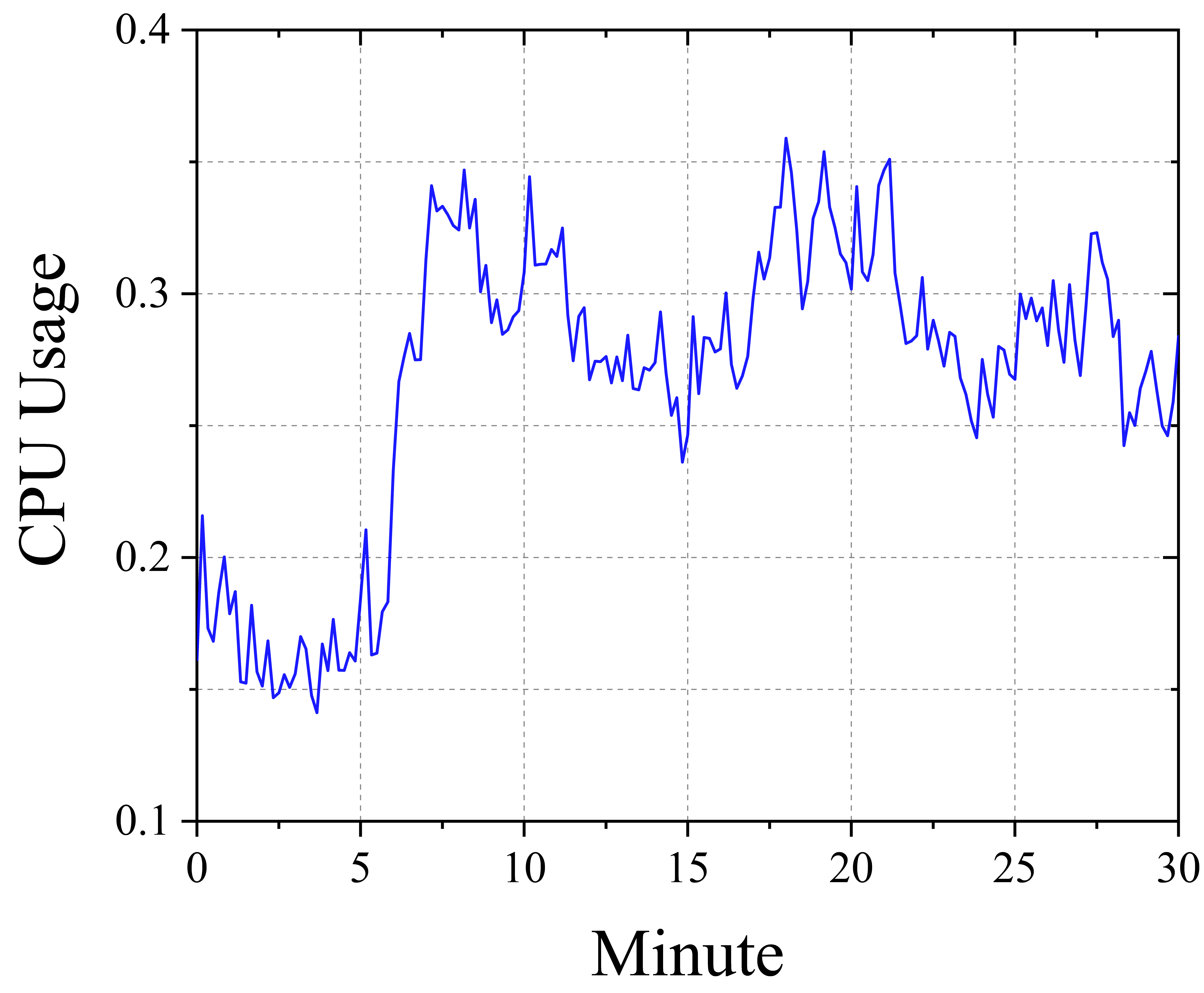}
	\end{subfigure}
	\caption{Alibaba (a) per-day workload (b) per-minute workload fluctuations}
	\label{fig:Alibaba_CPU}
\end{figure*}

\begin{figure*}[t]
	\centering
	\begin{subfigure}[b]{0.49\linewidth}
		\centering
		\includegraphics[height=4cm, width=0.99\linewidth]{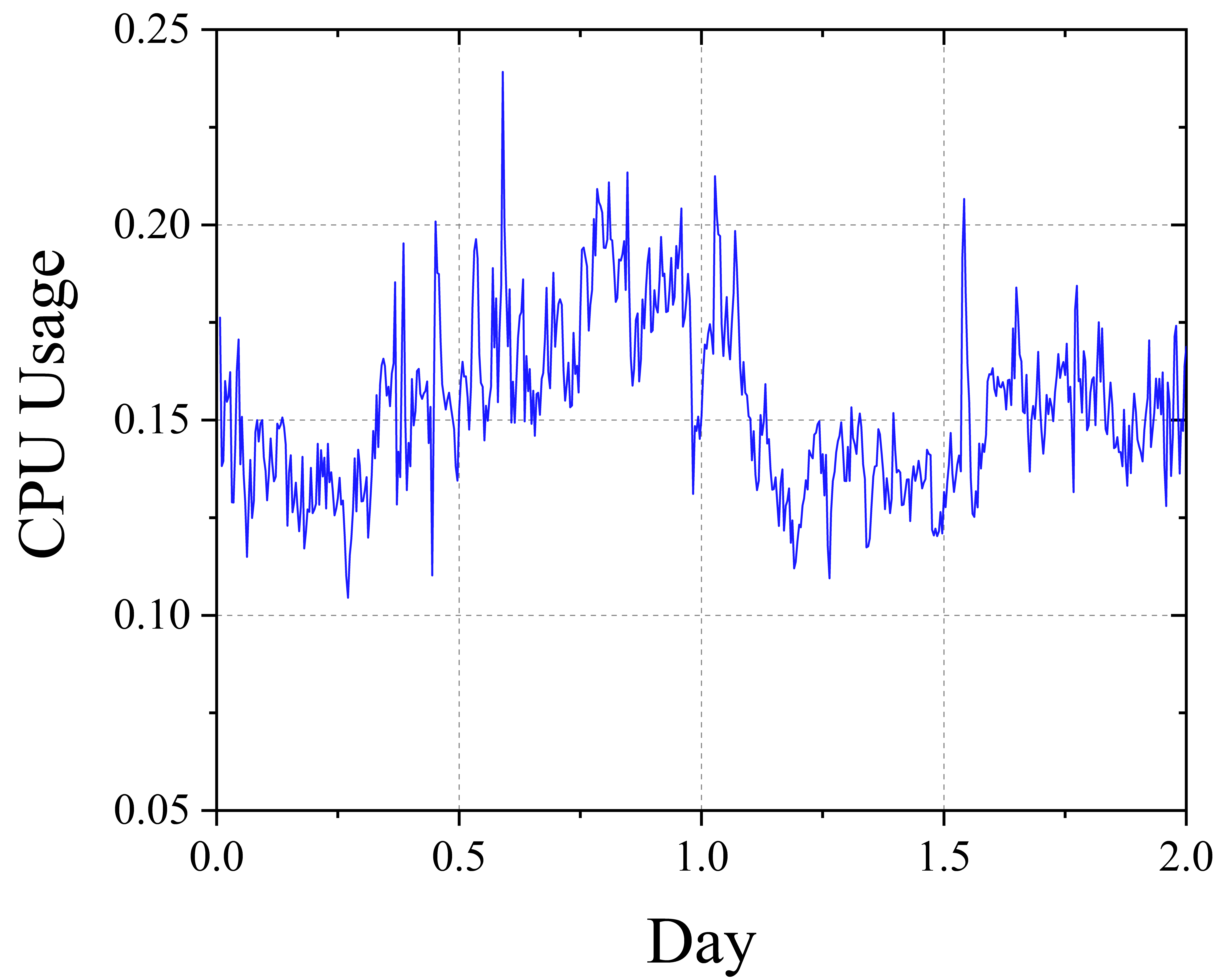}
\label{Google per-day workload}
	\end{subfigure}
	\begin{subfigure}[b]{0.49\linewidth}
		\centering
		\includegraphics[height=4cm, width=0.99\linewidth]{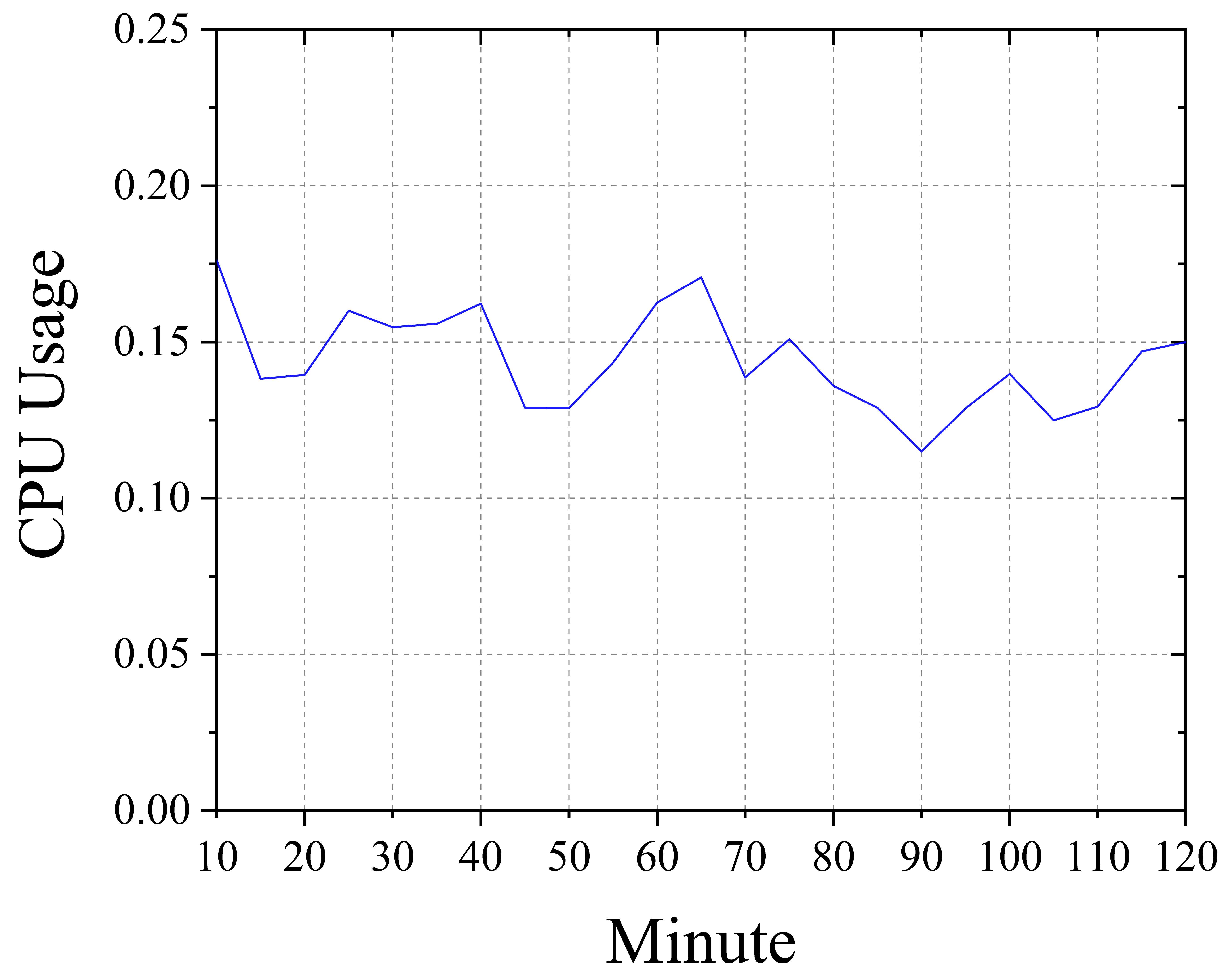}
\label{Google per-minute workload}
	\end{subfigure}
	\caption{Google (a) per-day workload (b) per-minute workload fluctuations}
	\label{fig:Google_CPU}
\end{figure*}

     Figs.~\ref{fig:Alibaba_CPU} and \ref{fig:Google_CPU} demonstrate the CPU usage in the datasets of Alibaba and Google cloud data centers, respectively. We have divided them into per-day and per-minute workloads fluctuations of machines so that we can see the fluctuations of CPU usage more clearly over time. We can notice that both the datasets show high variance and random features. 
	For the Alibaba dataset, we divide the dataset into the first 40,000 rows of data (59.5\%) and the rest, which are used to train and test the model. Similarly, we have divided the Google dataset in this way. \textcolor{black}{We selected about 72 hours of data from Google's dataset, and we used the first 49 hours (68.4\%) as the training set and the rest as the validation set.} For these two datasets, the number of training epoch is 200, the batch size is 72, the loss function is Huber, the optimizer is Adam, and the metric we use is mean square errors.

%
%
	

	 \subsection{Comparison with Unsupervised Learning-based Approach}
	In contrast to the supervised learning approach used by esDNN, the unsupervised learning approach can also be applied to high-dimensional problems such as multivariate time series forecasting, therefore, in this section, we evaluate our approach with unsupervised learning-based approach.
	
	Among the unsupervised learning approaches, Autoencoder is a representative one for efficient feature extraction and feature representation of high-dimensional data \cite{bourlard1988auto}. Currently, Autoencoder as well as Stacked Autoencoder, Sparse Autoencoder \cite{ng2011sparse}, and Denoising Autoencoder \cite{lu2013speech} are widely used in the research field. Autoencoder maps the input sample $x$ to the hidden layer by the encoder ($g$) and then maps it back to the original space by the decoder ($f$) to obtain the reconstructed sample. For the neural network-based autoencoder model, the encoder part compresses the data by reducing the number of neurons layer by layer, while the decoder part improves the number of neurons layer by layer based on the abstract representation of the data, and finally realizes the reconstruction of the input samples. The optimization objective is to optimize both the encoder and decoder by minimizing the Loss function. The optimization equation is shown as below:
	\begin{equation}
		f, g=\operatorname{min}_{f, g} Loss(x, f(g(x))).
	\end{equation}
	
\begin{figure*}[ht]
		\centering
		\includegraphics[width=0.5\linewidth]{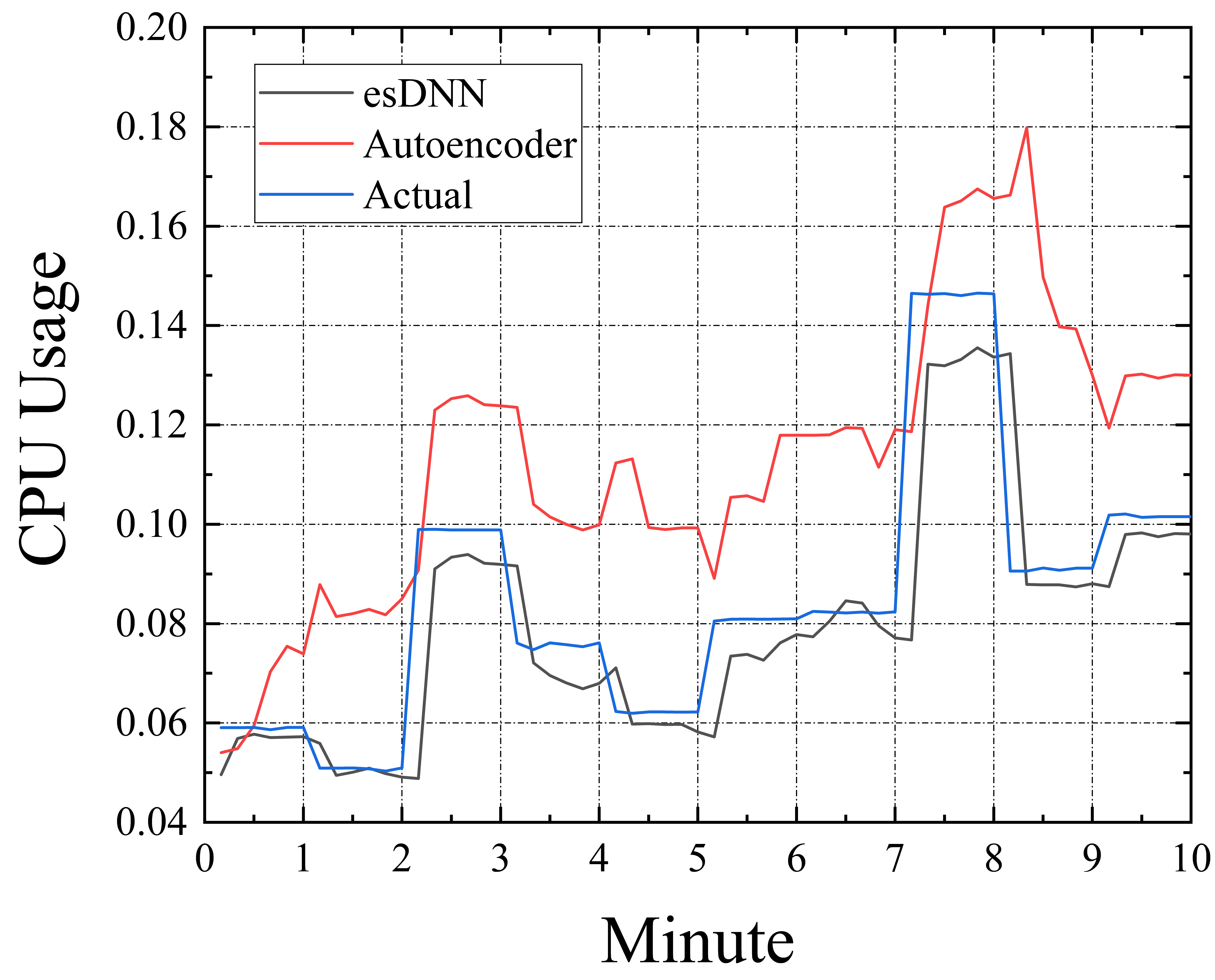}
		\caption[VarPerOptCom]{\color{black}Comparison of the prediction results of esDNN and Autoencoder}
		\label{fig:Autoencoder_compare}
	\end{figure*}

	The prediction results of esDNN and Autoencoder within 10 minutes are shown in Fig.~\ref{fig:Autoencoder_compare}, which demonstrates that Autoencoder has a good prediction result only at the beginning of the observed period, and it is significantly less accurate than esDNN. The reason can be that Autoencoder does not need to use the label of the sample in prediction, and it uses the input of the sample as both the input and output of the neural network. Although this can greatly improve the generality of the model, autoencoder is prone to be overfitting when the parameters of the neural network are complicated. Based on the results compared with unsupervised learning, supervised learning based approach has demonstrated better performance. In the following experiments, we evaluate the performance with other neural network-based approaches.  
	
	
	\subsection{Comparison with Neural Network-based Approaches}
We first start the evaluation with the Alibaba dataset. \color{black}To compare with our algorithm, we have selected several RNN-based deep learning algorithms, which have been applied for time series prediction, including RNN, Bi-LSTM \cite{bi2021integrated} and GRU. \color{black}We compare the prediction accuracy of these four algorithms and measure them by Mean Square Errors (MSE), which is represented as: 
		\begin{equation}
			MSE=\frac{1}{m} \sum_{i=1}^{m}\left(y_{actual}-{y}_{predict}\right)^{2},
		\label{MSE}
		\end{equation}	
where $m$ represents the timestamp, $y_{actual}$ is the actual value and $y_{predict}$ is the predictive value. The higher the MSE of the algorithm, the greater the gap between the predicted value and the actual value. In order to capture the changing trend of MSE in each period, we set four different time scales: second, minute, hour and day.

\begin{figure*}[ht!]
	\centering
	\begin{subfigure}[b]{0.49\linewidth}
		\centering
		\includegraphics[height=4cm, width=0.9\linewidth]{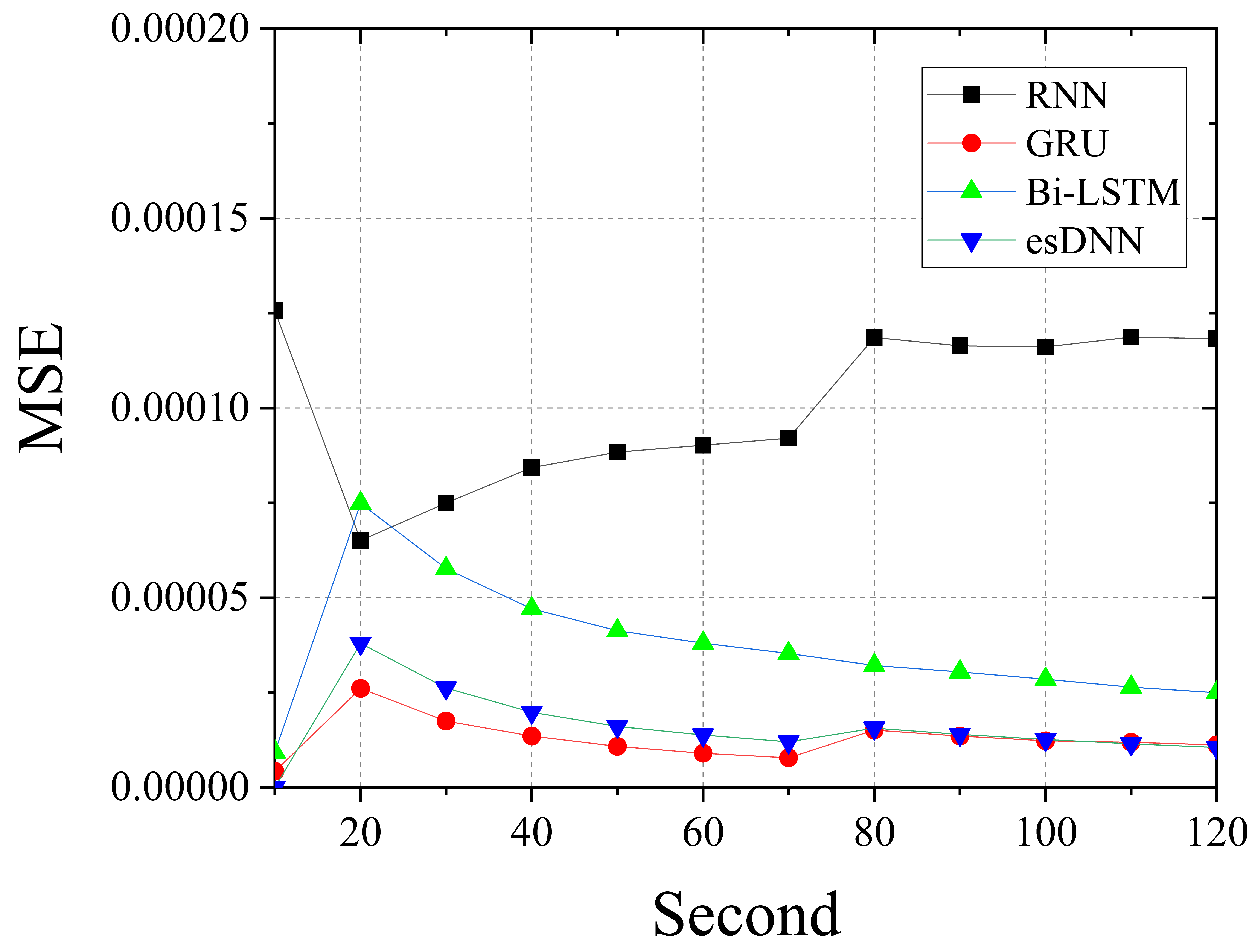}
\label{Alibaba second level prediction}
	\end{subfigure}
	\begin{subfigure}[b]{0.49\linewidth}
		\centering
		\includegraphics[height=4cm, width=0.9\linewidth]{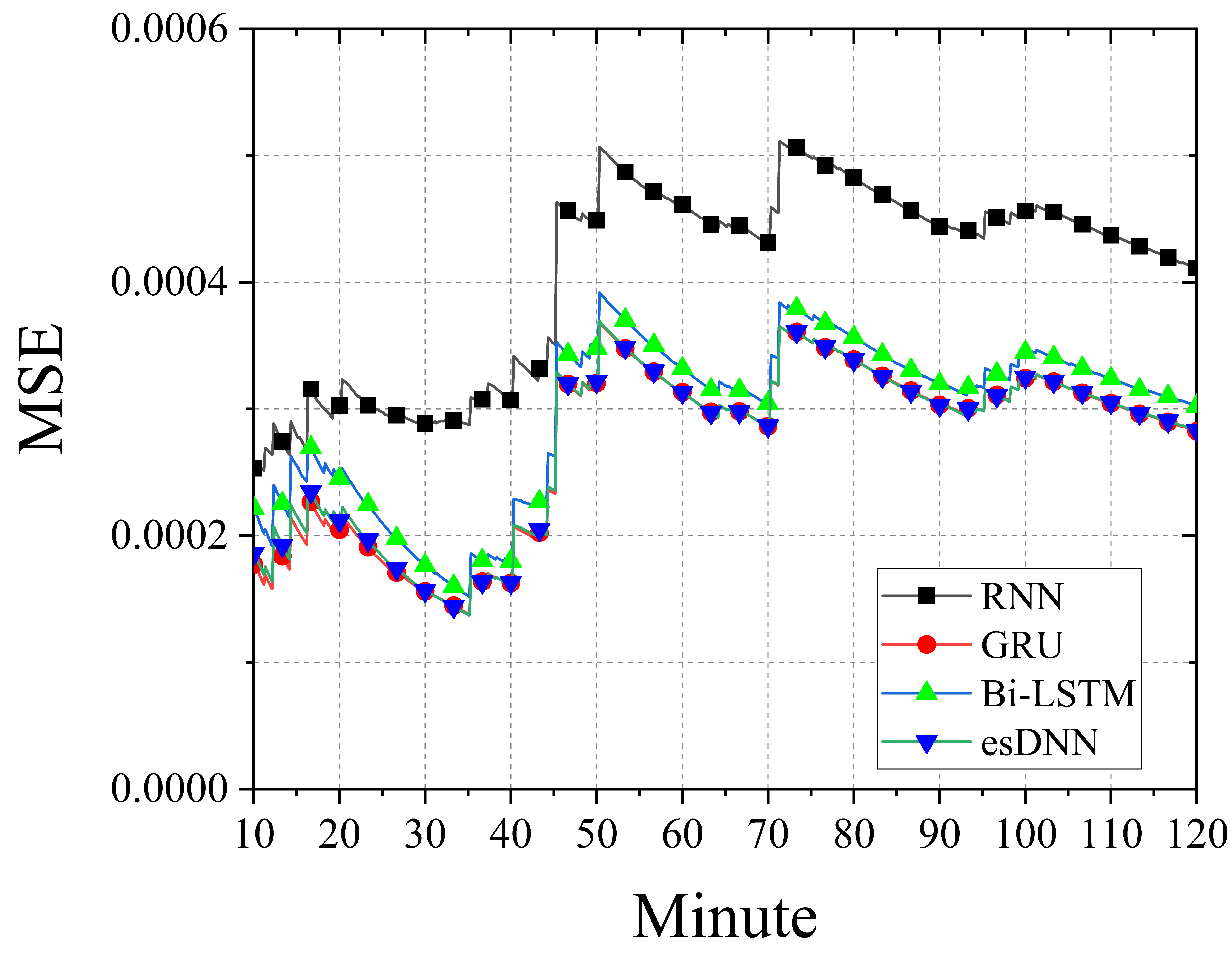}
\label{Alibaba minute level prediction}
	\end{subfigure}
    	\begin{subfigure}[b]{0.49\linewidth}
		\centering
		\includegraphics[height=4cm, width=0.9\linewidth]{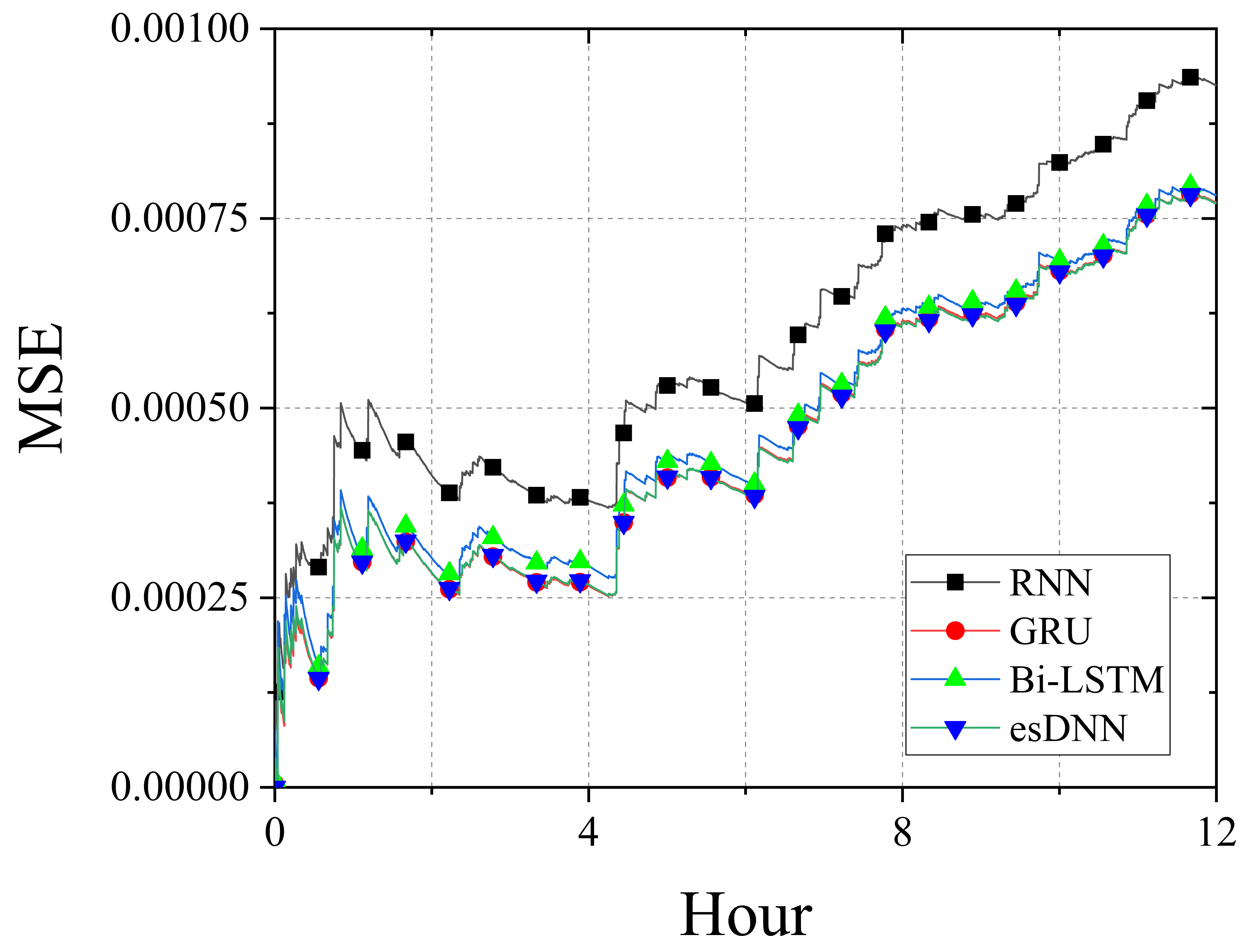}
\label{Alibaba hour level prediction}
	\end{subfigure}
    	\begin{subfigure}[b]{0.49\linewidth}
		\centering
		\includegraphics[height=4cm, width=0.9\linewidth]{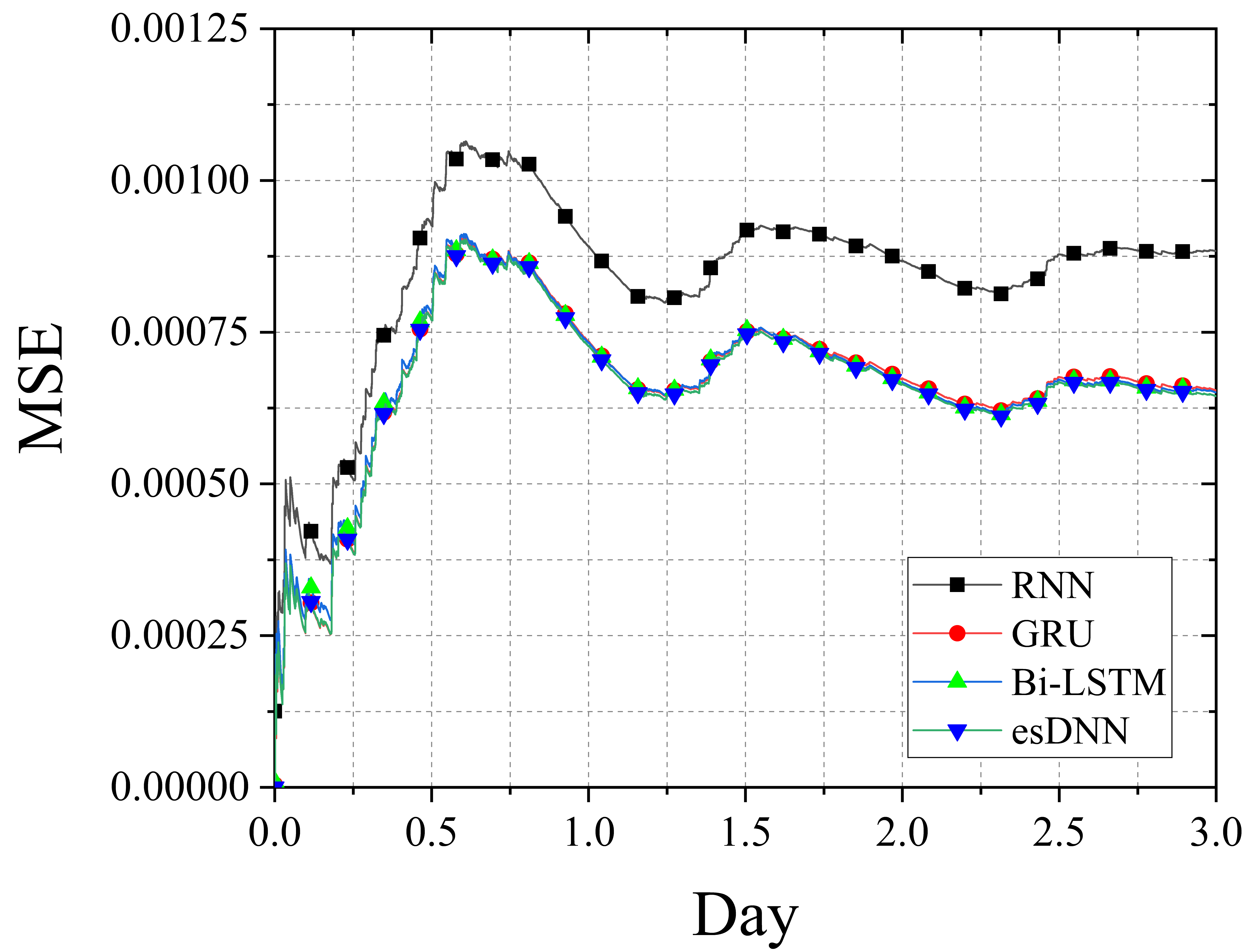}
\label{Alibaba day level prediction}
	\end{subfigure}
	\caption{\color{black}Prediction accuracy (MSE) of four different RNN-based methods based on the Alibaba dataset with different prediction lengths}
	\label{fig:Alibaba MSE}
\end{figure*}

\color{black}Fig.~\ref{fig:Alibaba MSE} shows the MSE fluctuation of these four RNN-based methods based on the Alibaba dataset with various prediction lengths. In general, all the MSE curves follow the same trend, which shows that the MSE value first increases until it reaches a peak, after that, the curves will remain at a relatively stable value. For the second-level prediction, apart from keeping RNN at a relatively high value, there are just subtle differences between Bi-LSTM, GRU and esDNN. With the increase of the prediction length, all these curves are maintained at relatively stable values. But for the day-level prediction, there is no significant difference in MSE value between Bi-LSTM, GRU and esDNN. \color{black}The main reason is that the RNN and Bi-LSTM models are designed to process time-series data and can perform well in representing the nonlinear relationship between data and time. However, the drawback of RNN and Bi-LSTM is that their gradient can disappear or explode, especially for the long time-series data during the data training process. The GRU can alleviate the side effects of gradient disappearance that usually happens in Bi-LSTM and RNN. Therefore, the results of GRU-based approach can maintain relatively more stable values compared with RNN and Bi-LSTM.   \color{black}
	
	In order to brighten the differences between them, we choose to use the Cumulative Distribution Function (CDF) method to measure them, which is the integral of the probability density function. For discrete variables, it can represent the sum of the probability of occurrence of all values less than or equal to $x$, which is formulated as:
		\begin{equation}
			F_{X}(x)=\mathrm{P}(X \leq x).
			\label{prob}
		\end{equation}
	
    
\begin{figure*}[t]
	\centering
	\label{fig:CDF MSE}
	\begin{subfigure}[b]{0.48\linewidth}
		\centering
		\includegraphics[height=4cm, width=0.9\linewidth]{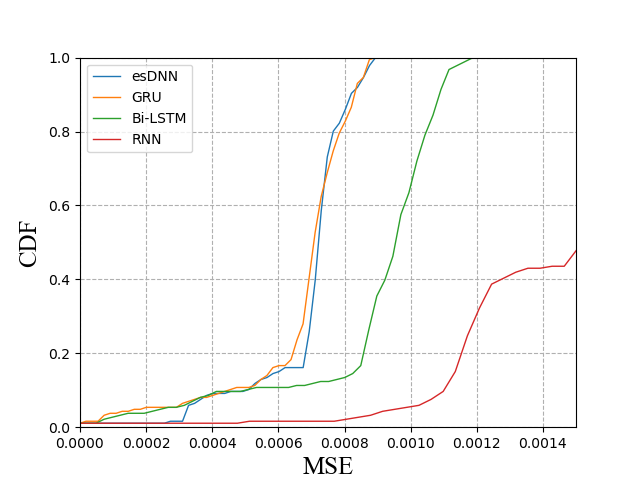}
				\caption{\color{black}CDF of MSE with Alibaba Trace}
	\end{subfigure}
	\begin{subfigure}[b]{0.48\linewidth}
		\centering
		\includegraphics[height=4cm, width=0.9\linewidth]{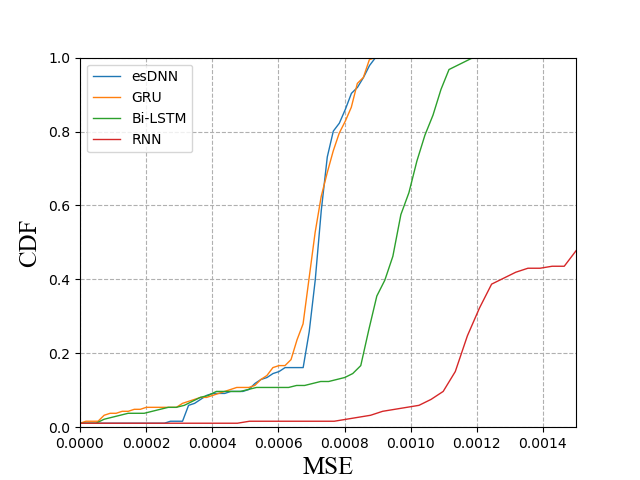}
				\caption{\color{black}CDF of MSE with Google Trace}
	\end{subfigure}
	\caption{\color{black}CDF comparison of MSE curve}
	\label{fig:CDF}
\end{figure*}

\color{black}	
Fig.~\ref{fig:CDF} shows the CDF of MSE based on these four RNN-based methods, these almost overlapping curves in Fig.~\ref{fig:Alibaba MSE} can be distinguished more apparently by the difference in CDF values. Since RNN is quite different from the other three methods, and its effect is the worst one, therefore, we focus on the discussions on the other three algorithms. Except for the RNN method, we can clearly see that when the value of MSE is between 0.006 and 0.008, the value of CDF rises very quickly, which shows that the MSE values of these three methods are concentrated. Meanwhile, esDNN rises significantly faster than Bi-LSTM and GRU, we can see that for any MSE in this time period, the CDF value of esDNN always remains at the highest value. Although the curves are very close, the difference in value between them can still be easily identified. It means that the overall MSE value of esDNN is smaller than the values of Bi-LSTM and GRU.
\color{black}
	
    
\begin{figure*}[t]
	\centering
	\begin{subfigure}[b]{0.48\linewidth}
		\centering
		\includegraphics[height=4cm, width=0.9\linewidth]{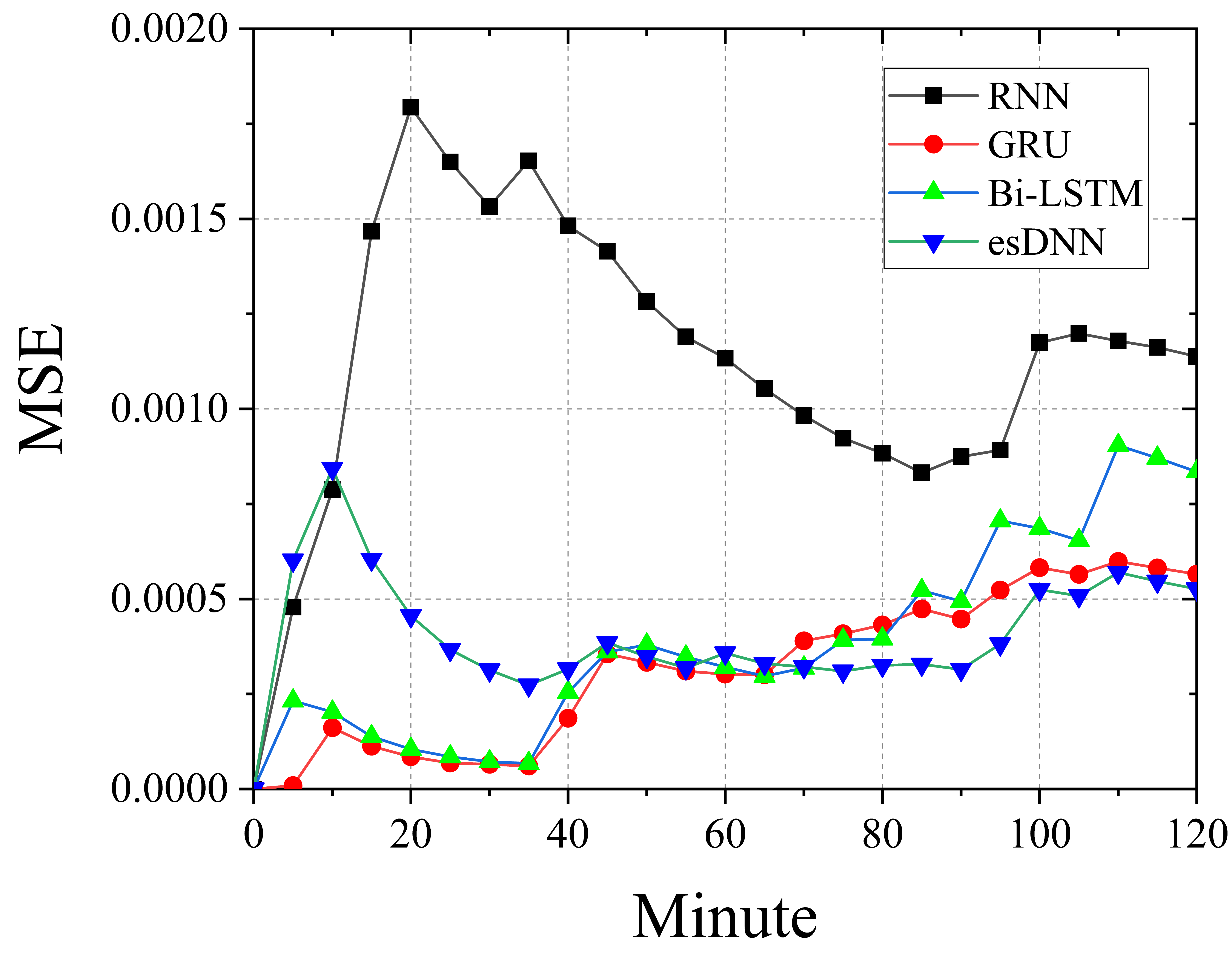}
\label{Google minute level prediction}
	\end{subfigure}
	\begin{subfigure}[b]{0.48\linewidth}
		\centering
		\includegraphics[height=4cm, width=0.9\linewidth]{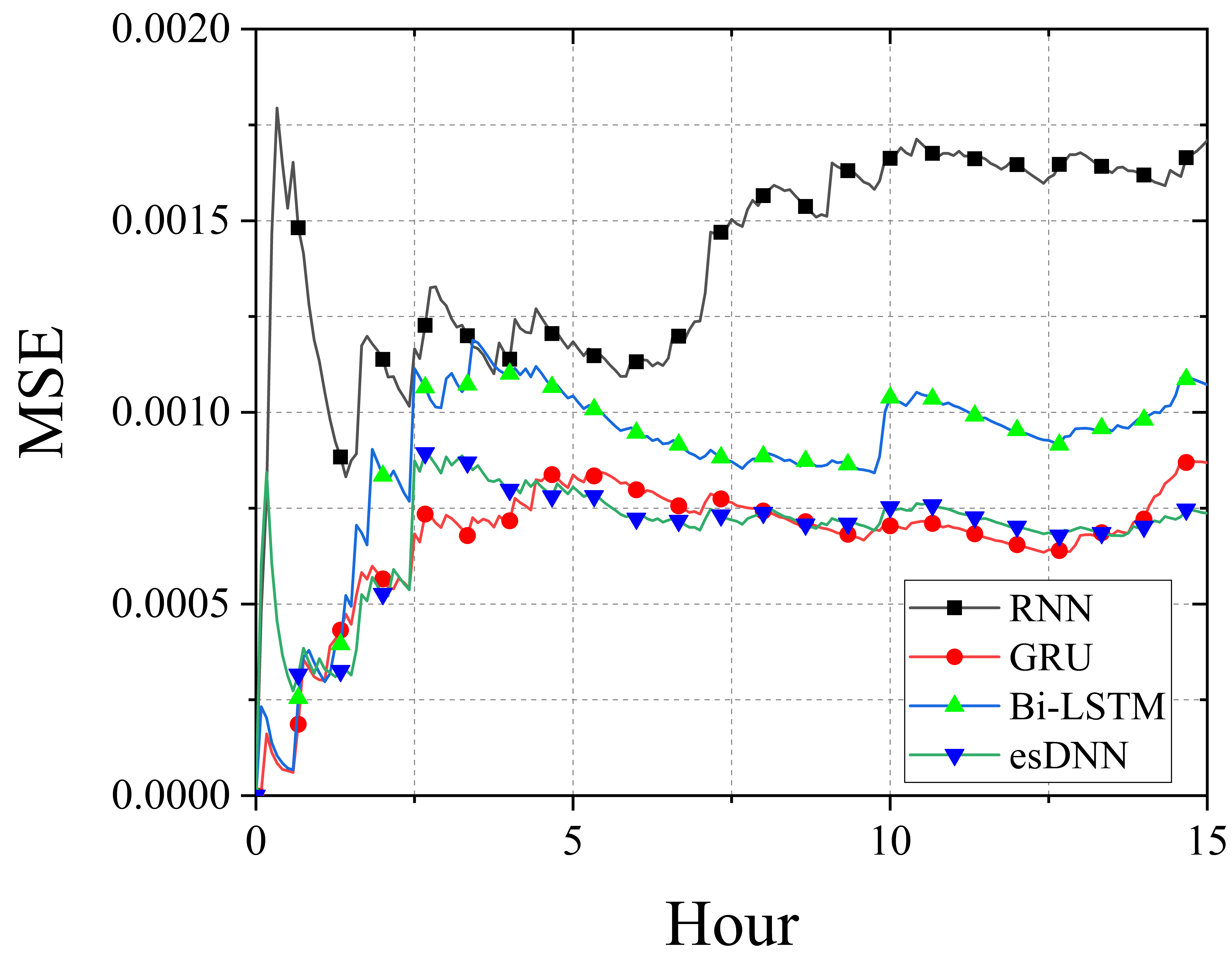}
\label{Google hour level prediction}
	\end{subfigure}
	\caption{\color{black}Prediction accuracy (MSE) of four different RNN methods based on the Google dataset}
	\label{fig:Google MSE}
\end{figure*}

For Google's dataset, we also use the RNN-based methods as baselines for esDNN. Compared with the Alibaba dataset, we utilize less data from Google's dataset, therefore, we show the prediction length with minute-level and hour-level. \color{black}Fig.~\ref{fig:Google MSE} shows the MSE fluctuation of the four methods based on the Google dataset. For the minute-level prediction, all these methods have little difference between each other except RNN. When we focus on the hour-level prediction, the trend of these methods is stable after a short growth, which is consistent with the results of the Alibaba dataset. We can also notice that RNN always maintains at a high level. Compared with GRU and esDNN, Bi-LSTM has a higher MSE. For esDNN and GRU, the MSE of these two are quite close, where the esDNN can achieve a more stable trend, while GRU fluctuates more dynamically. To conclude, we can notice that the prediction result of esDNN is better than GRU, since the MSE value of esDNN is smaller than GRU. \color{black}
	
\begin{figure*}[t]
	\centering
	\begin{subfigure}[b]{0.48\linewidth}
		\centering
		\includegraphics[height=4cm, width=0.9\linewidth]{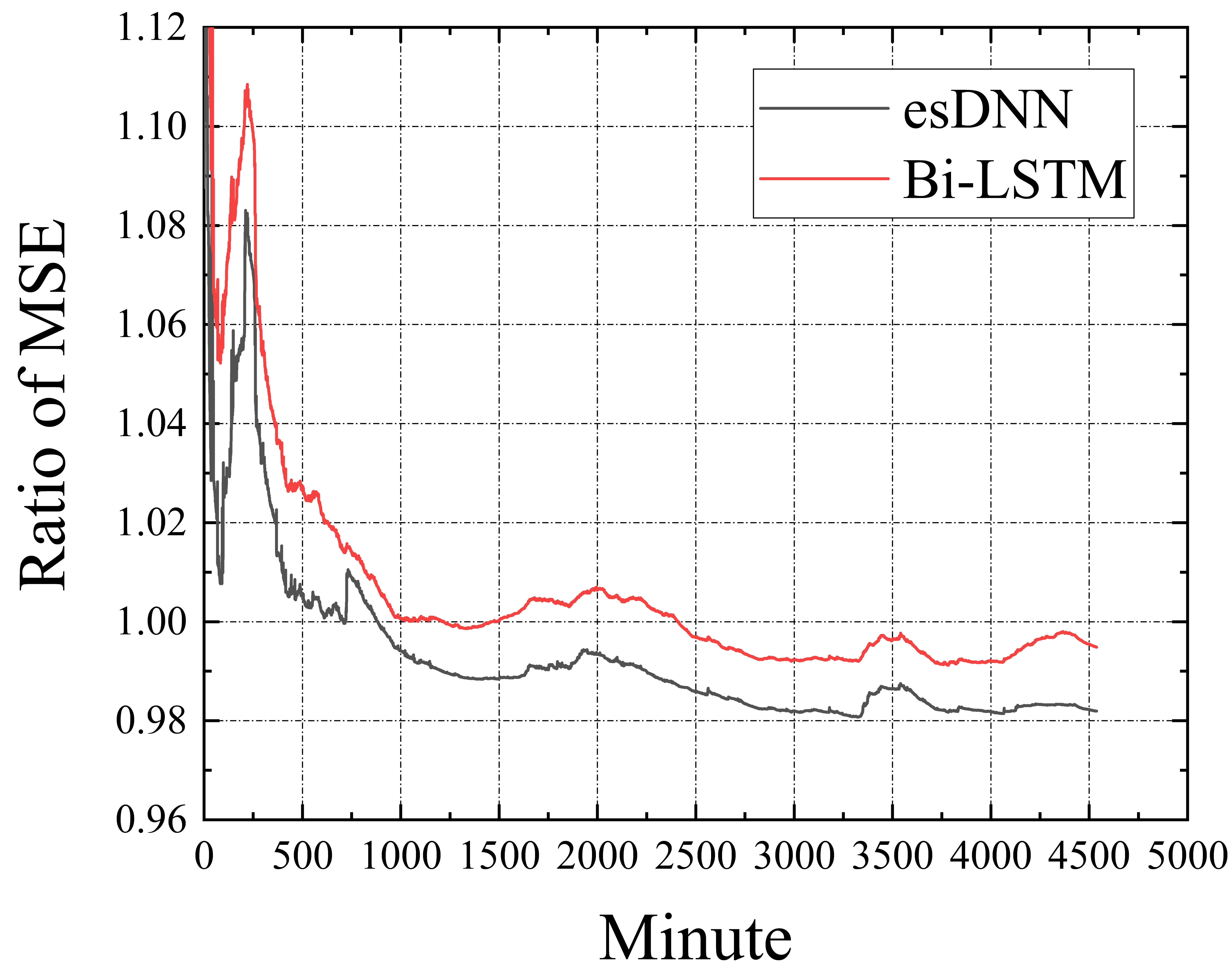}
			\caption{\color{black}Ratio of MSE with Alibaba Trace}
\label{Alibaba Ratio of MSE}
	\end{subfigure}
	\begin{subfigure}[b]{0.48\linewidth}
		\centering
		\includegraphics[height=4cm, width=0.9\linewidth]{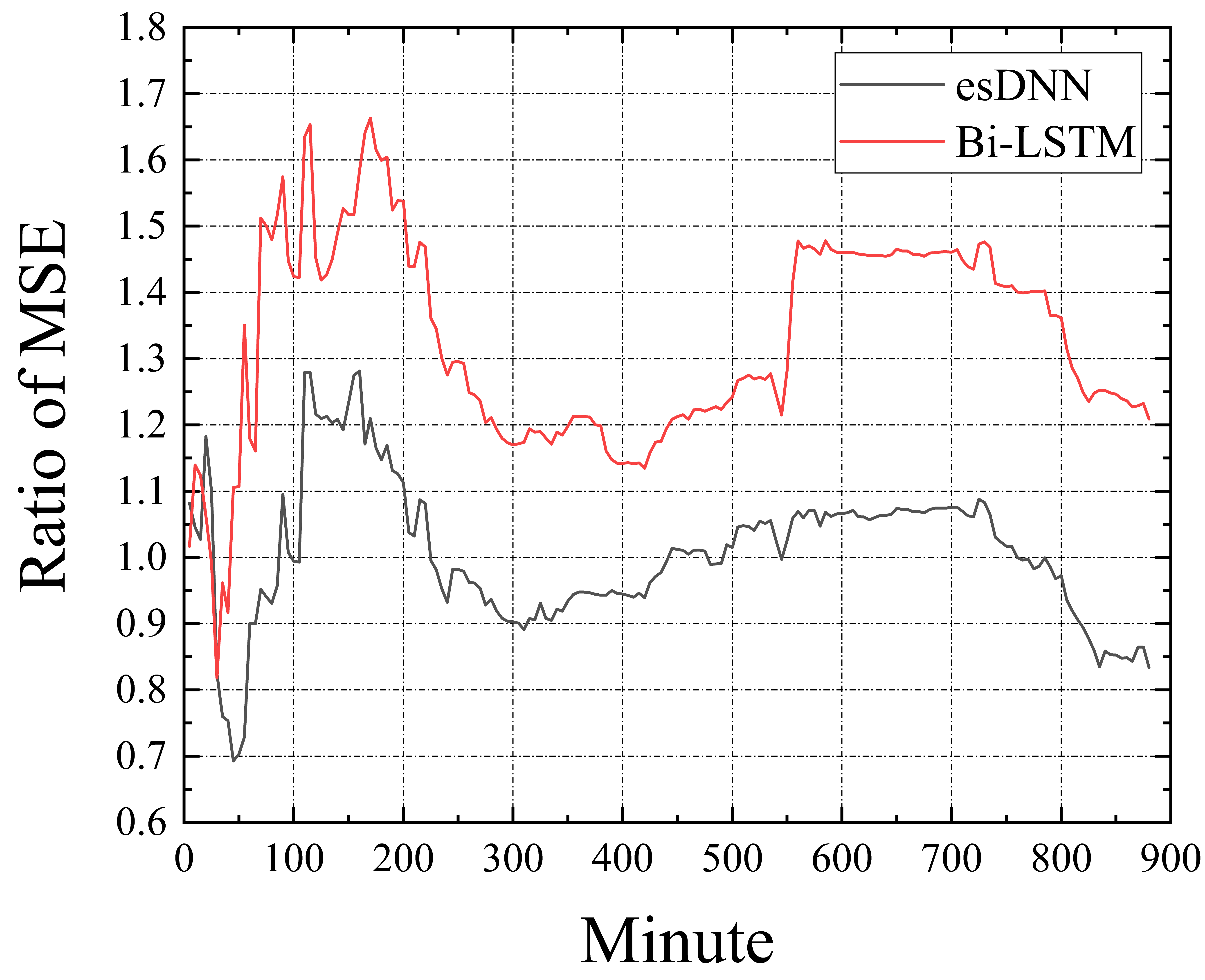}
			\caption{\color{black}Ratio of MSE with Google Trace}
\label{Google Ratio of MSE}
	\end{subfigure}
	\caption{\color{black}The ratio of MSE compared with GRU for Alibaba and Google Traces}
	\label{fig:GRU_MSE}
\end{figure*}

\textcolor{black}{We notice that esDNN is very close to GRU's results over a long sequence of time while the trend of RNN is much worse than the other three approaches. Thus we choose to compare the MSE values without RNN. We analyze the MSE values from Alibaba and Google dataset separately, as shown in Fig.~\ref{fig:GRU_MSE}, where the GRU is set as the baseline, and the MSE values of esDNN and Bi-LSTM are divided by the MSE values of GRU. The value less than 1.0 represents better performance than GRU, and vice versa. Although GRU performs better prediction results than esDNN in a short period of time, esDNN can maintain better accuracy and stability in the long run.}

\begin{figure*}[t]
	\centering
	\begin{subfigure}[b]{0.48\linewidth}
		\centering
		\includegraphics[height=4cm, width=0.9\linewidth]{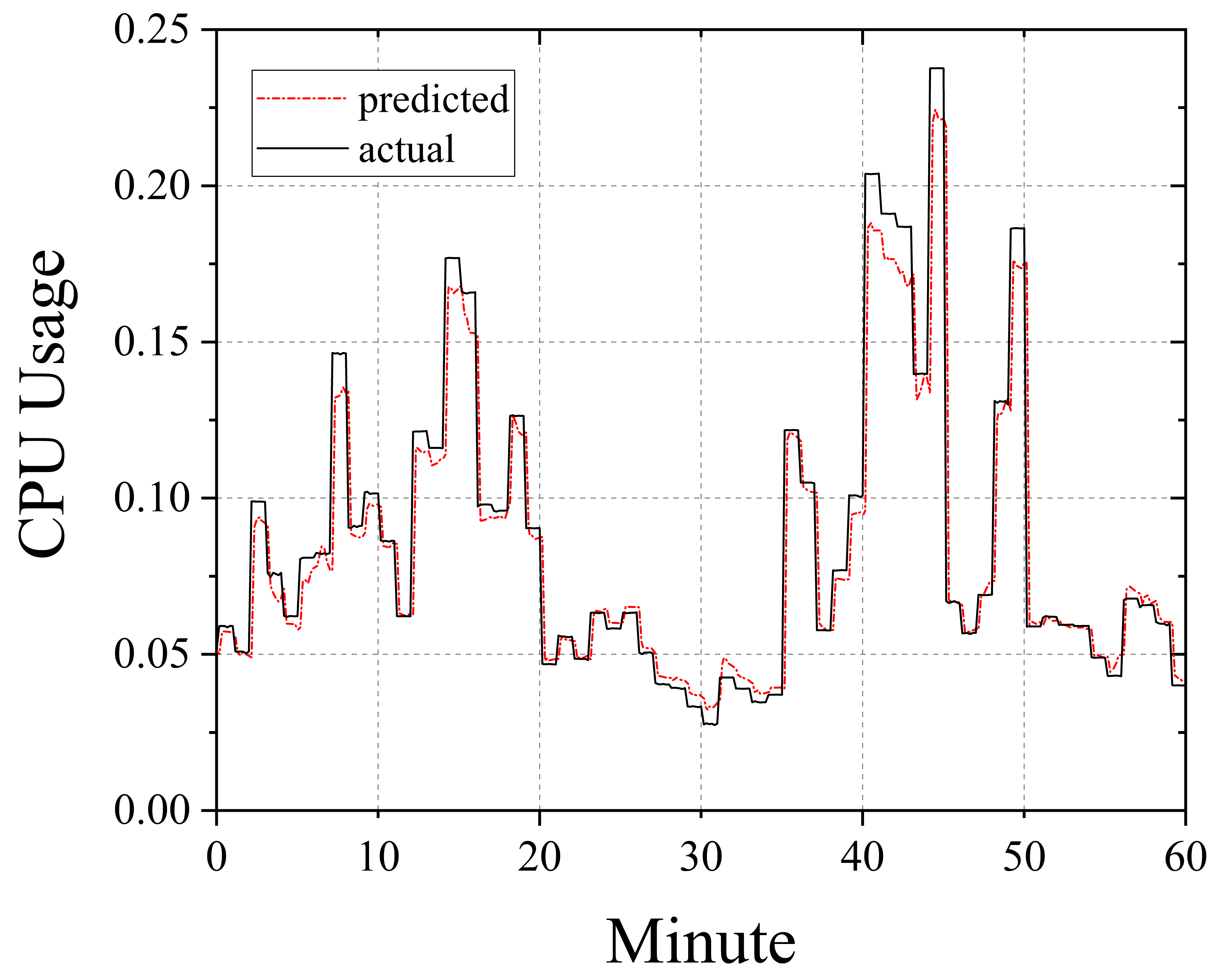}
\label{fig:Alibaba esDNN minute prediction}
	\end{subfigure}
	\begin{subfigure}[b]{0.48\linewidth}
		\centering
		\includegraphics[height=4cm, width=0.9\linewidth]{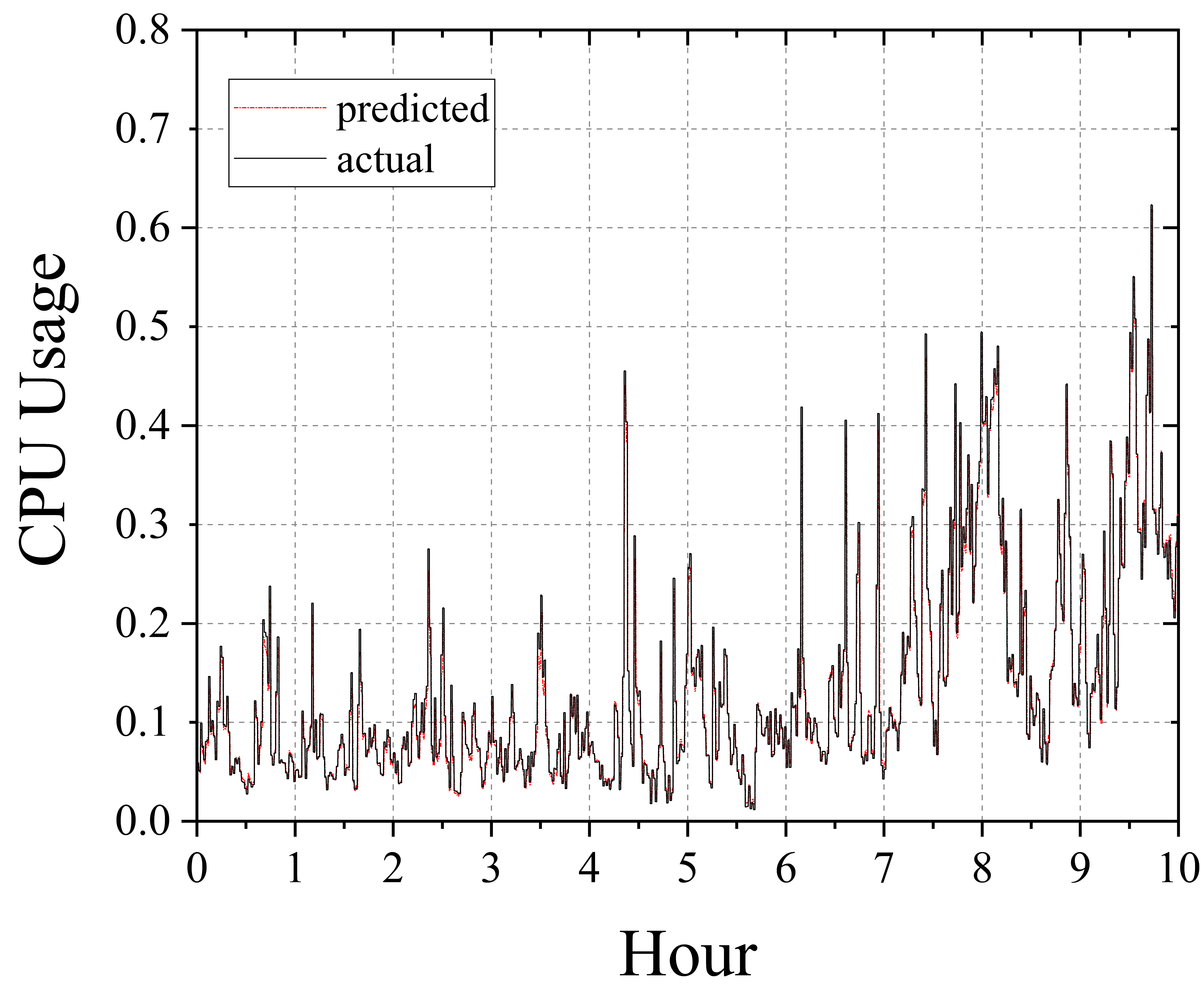}
\label{fig:Alibaba esDNN hour prediction}
	\end{subfigure}
	\caption{\color{black}Prediction performance of the esDNN compared with actual Alibaba data}
	\label{fig:Alibaba esDNN}
\end{figure*}

\begin{figure*}[t]
	\centering
	\begin{subfigure}[b]{0.48\linewidth}
		\centering
		\includegraphics[height=4cm, width=0.9\linewidth]{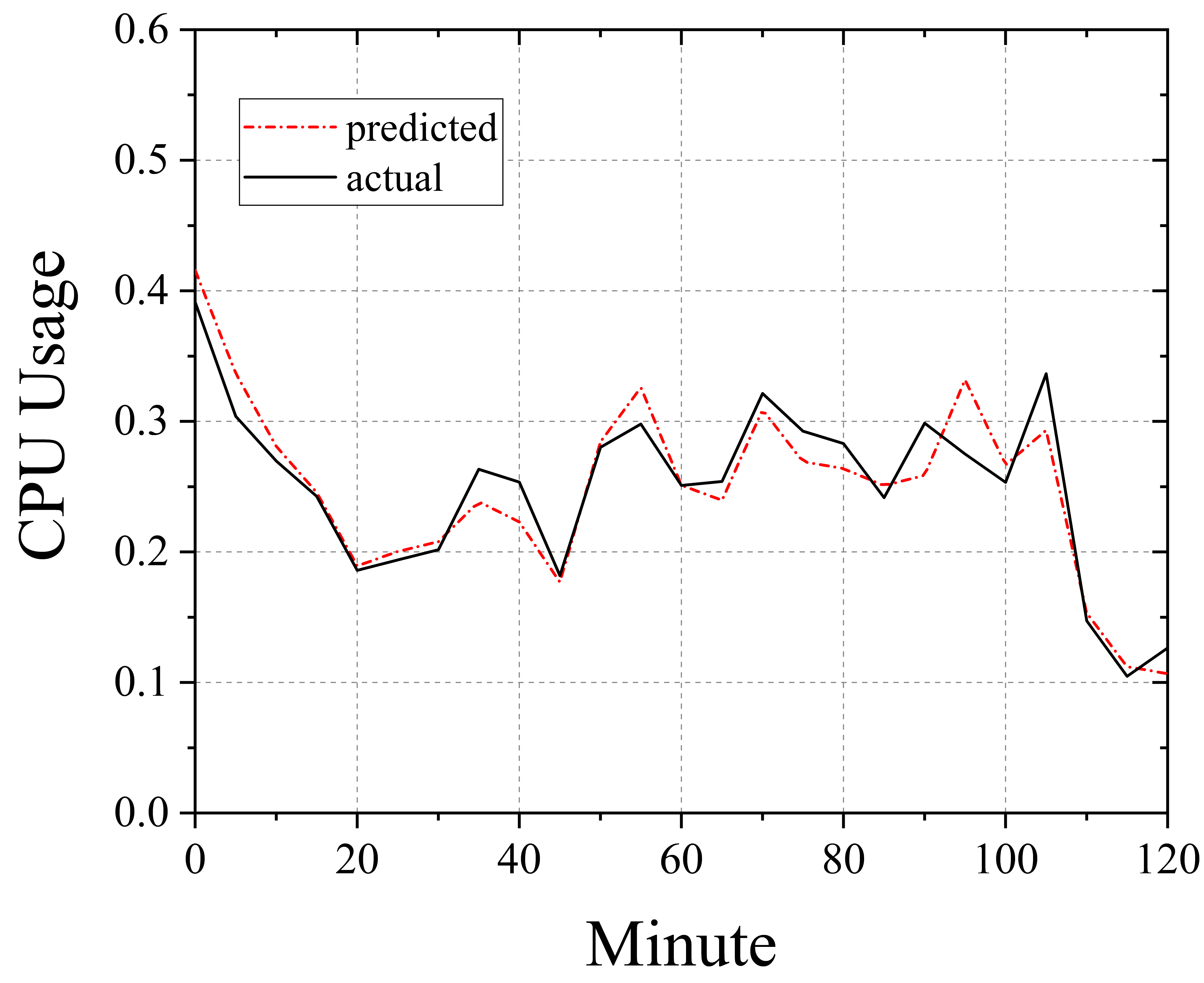}
\label{fig:Google esDNN minute prediction}
	\end{subfigure}
	\begin{subfigure}[b]{0.48\linewidth}
		\centering
		\includegraphics[height=4cm, width=0.9\linewidth]{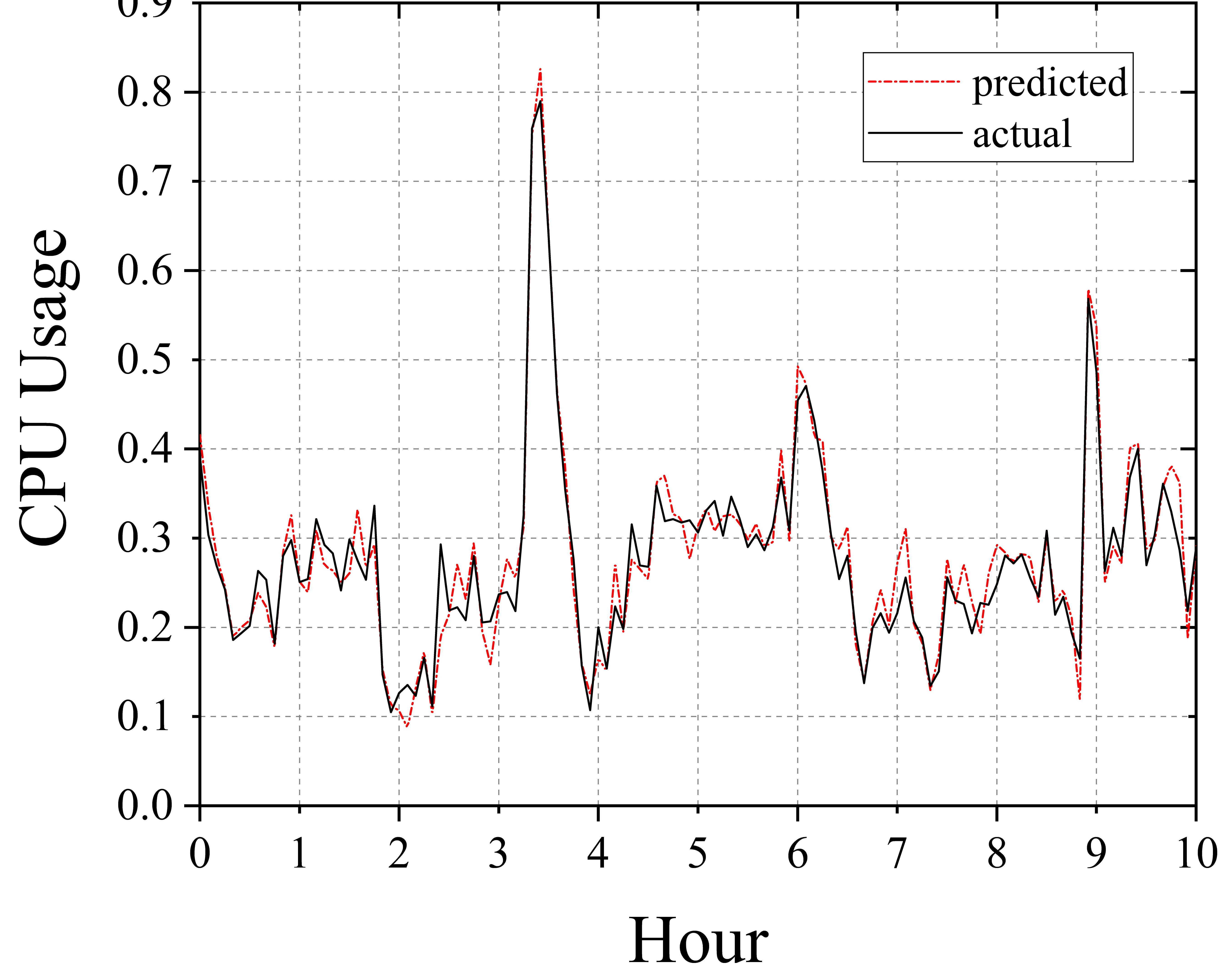}
\label{fig:Google esDNN hour prediction}
	\end{subfigure}
	\caption{\color{black}Prediction performance of the esDNN compared with actual Google data}
	\label{fig:Google esDNN}
\end{figure*}

Next, we evaluate the difference between the predicted value and the actual value of esDNN. Figs.~\ref{fig:Alibaba esDNN} and \ref{fig:Google esDNN} show the CPU usage curves, so that we can see the difference between the predicted value and the actual value. For the analysis of the Alibaba dataset, we can observe that for the minute level prediction, though there are some large differences between consecutive values, esDNN can still give relatively accurate prediction results. For the hour-level prediction, on the whole, the predicted value is very close to the actual value because their curves are almost fit, and only a small part of the predicted curve is different from the actual value. For hour-level prediction based on Google cluster data, esDNN can still accurately predict the trend of CPU usage. 

\color{black} \color{black}In order to identify the difference between them more intuitively, we summarize the performance of these algorithms as listed in Table \ref{tab:Alibaba table} and Table \ref{tab:Google table}. \color{black}Apart from MSE, we also compare Root Mean Square Error (RMSE) and Mean Absolute Percentage Error (MAPE) that have been widely used to evaluate prediction performance. The esDNN approach can achieve the lowest MSE values compared with other baselines when the prediction length is longer, which is more difficult to be predicted. For the Google dataset, our approach can also achieve the lowest RMSE with a longer prediction length. The reason is that our proposed prediction approach based on revised GRU and CNN not only captures the periodical features inherent in the data, but also significantly reduces the impact of resource variations on prediction results.   \color{black}

\begin{table}[]
	\caption{\color{black}MSE, RMSE and MAPE comparison with Alibaba dataset}
	\label{tab:Alibaba table}
	\resizebox{\textwidth}{!}{%
		\begin{tabular}{|c|c|c|c|c|c|c|c|c|c|c|c|c|}
			\hline
{\color{black} } & \multicolumn{3}{c|}{{\color{black} \textbf{RNN}}} & \multicolumn{3}{c|}{{\color{black} \textbf{GRU}}} & \multicolumn{3}{c|}{{\color{black} \textbf{Bi-LSTM}}} & \multicolumn{3}{c|}{{\color{black} \textbf{esDNN}}} \\ \cline{2-13} 
\multirow{-2}{*}{{\color{black} \textbf{\begin{tabular}[c]{@{}c@{}}Prediction\\ length\end{tabular}}}} & {\color{black} \textbf{MSE}} & {\color{black} \textbf{RMSE}} & {\color{black} \textbf{MAPE}} & {\color{black} \textbf{MSE}} & {\color{black} \textbf{RMSE}} & {\color{black} \textbf{MAPE}} & {\color{black} \textbf{MSE}} & {\color{black} \textbf{RMSE}} & {\color{black} \textbf{MAPE}} & {\color{black} \textbf{MSE}} & {\color{black} \textbf{RMSE}} & {\color{black} \textbf{MAPE}} \\ \hline
{\color{black} 10s} & {\color{black} 1.25E-04} & {\color{black} 0.0112} & {\color{black} 0.2175} & {\color{black} 4.27E-06} & {\color{black} 0.0021} & {\color{black} 0.0401} & {\color{black} 9.33E-06} & {\color{black} 0.0031} & {\color{black} 0.1596} & {\color{black} \textbf{1.59E-07}} & {\color{black} \textbf{0.0004}} & {\color{black} \textbf{0.0077}} \\ \hline
{\color{black} 30s} & {\color{black} 7.50E-05} & {\color{black} 0.0087} & {\color{black} 0.1396} & {\color{black} \textbf{1.75E-05}} & {\color{black} \textbf{0.0042}} & {\color{black} \textbf{0.0541}} & {\color{black} 5.77E-05} & {\color{black} 0.0076} & {\color{black} 0.1086} & {\color{black} 3.13E-05} & {\color{black} 0.0056} & {\color{black} 0.0681} \\ \hline
{\color{black} 1 min} & {\color{black} 9.02E-05} & {\color{black} 0.0095} & {\color{black} 0.1568} & {\color{black} \textbf{9.02E-06}} & {\color{black} \textbf{0.0030}} & {\color{black} \textbf{0.0325}} & {\color{black} 3.80E-05} & {\color{black} 0.0062} & {\color{black} 0.0891} & {\color{black} 1.71E-05} & {\color{black} 0.0041} & {\color{black} 0.0480} \\ \hline
{\color{black} 30 min} & {\color{black} 2.93E-04} & {\color{black} 0.0170} & {\color{black} 0.2024} & {\color{black} \textbf{1.71E-04}} & {\color{black} \textbf{0.0125}} & {\color{black} 0.0772} & {\color{black} 2.00E-04} & {\color{black} 0.0142} & {\color{black} 0.0978} & {\color{black} 1.84E-04} & {\color{black} 0.0128} & {\color{black} \textbf{0.0766}} \\ \hline
{\color{black} 1h} & {\color{black} 4.61E-04} & {\color{black} 0.0215} & {\color{black} 0.2343} & {\color{black} \textbf{3.13E-04}} & {\color{black} \textbf{0.0177}} & {\color{black} 0.0936} & {\color{black} 3.31E-04} & {\color{black} 0.0182} & {\color{black} 0.1053} & {\color{black} 3.20E-04} & {\color{black} 0.0179} & {\color{black} \textbf{0.0891}} \\ \hline
{\color{black} 6h} & {\color{black} 5.07E-04} & {\color{black} 0.0225} & {\color{black} 0.2237} & {\color{black} \textbf{3.88E-04}} & {\color{black} \textbf{0.0197}} & {\color{black} 0.1050} & {\color{black} 4.03E-04} & {\color{black} 0.0200} & {\color{black} 0.\textbf{0891}} & {\color{black} 3.96E-04} & {\color{black} 0.0199} & {\color{black} 0.0990} \\ \hline
{\color{black} 1 day} & {\color{black} 8.91E-04} & {\color{black} 0.0293} & {\color{black} 0.2387} & {\color{black} 7.34E-04} & {\color{black} 0.0271} & {\color{black} 0.1260} & {\color{black} 7.32E-04} & {\color{black} 0.0271} & {\color{black} \textbf{0.0898}} & {\color{black} \textbf{7.25E-04}} & {\color{black} \textbf{0.0269}} & {\color{black} 0.1080} \\ \hline
{\color{black} 2 days} & {\color{black} 8.68E-04} & {\color{black} 0.0295} & {\color{black} 0.2058} & {\color{black} 6.74E-04} & {\color{black} 0.0260} & {\color{black} 0.1050} & {\color{black} 6.66E-04} & {\color{black} 0.0258} & {\color{black} \textbf{0.0872}} & {\color{black} \textbf{6.62E-04}} & {\color{black} \textbf{0.0257}} & {\color{black} 0.0901} \\ \hline
{\color{black} 3 days} & {\color{black} 8.83E-04} & {\color{black} 0.0297} & {\color{black} 0.1973} & {\color{black} 6.54E-04} & {\color{black} 0.0256} & {\color{black} 0.0978} & {\color{black} 6.50E-04} & {\color{black} 0.0255} & {\color{black} 0.0857} & {\color{black} \textbf{6.43E-04}} & {\color{black} \textbf{0.0254}} & {\color{black} \textbf{0.0842}} \\ \hline
\end{tabular}%
}
\end{table}

\color{black}


    \begin{table}[]
    	\caption{\color{black}MSE, RMSE and MAPE comparison with Google dataset}
    	\label{tab:Google table}
    	\resizebox{\textwidth}{!}{%
    		\begin{tabular}{|c|c|c|c|c|c|c|c|c|c|c|c|c|}
    			\hline
  {\color{black} } & \multicolumn{3}{c|}{{\color{black} \textbf{RNN}}} & \multicolumn{3}{c|}{{\color{black} \textbf{GRU}}} & \multicolumn{3}{c|}{{\color{black} \textbf{Bi-LSTM}}} & \multicolumn{3}{c|}{{\color{black} \textbf{esDNN}}} \\ \cline{2-13} 
  \multirow{-2}{*}{{\color{black} \textbf{\begin{tabular}[c]{@{}c@{}}Prediction\\ length\end{tabular}}}} & {\color{black} \textbf{MSE}} & {\color{black} \textbf{RMSE}} & {\color{black} \textbf{MAPE}} & {\color{black} \textbf{MSE}} & {\color{black} \textbf{RMSE}} & {\color{black} \textbf{MAPE}} & {\color{black} \textbf{MSE}} & {\color{black} \textbf{RMSE}} & {\color{black} \textbf{MAPE}} & {\color{black} \textbf{MSE}} & {\color{black} \textbf{RMSE}} & {\color{black} \textbf{MAPE}} \\ \hline
  {\color{black} 30min} & {\color{black} 0.00153232} & {\color{black} 0.0391} & {\color{black} 0.1651} & {\color{black} \textbf{6.53E-05}} & {\color{black} \textbf{0.0081}} & {\color{black} 0.0228} & {\color{black} 7.22E-05} & {\color{black} 0.0085} & {\color{black} \textbf{0.0209}} & {\color{black} 0.00031276} & {\color{black} 0.0177} & {\color{black} 0.0439} \\ \hline
  {\color{black} 1h} & {\color{black} 0.0011334} & {\color{black} 0.0337} & {\color{black} 0.1201} & {\color{black} \textbf{0.00030235}} & {\color{black} \textbf{0.0174}} & {\color{black} \textbf{0.0506}} & {\color{black} 0.00032074} & {\color{black} 0.0179} & {\color{black} 0.0567} & {\color{black} 0.0003576} & {\color{black} 0.0189} & {\color{black} 0.0507} \\ \hline
  {\color{black} 2h} & {\color{black} 0.00113796} & {\color{black} 0.0337} & {\color{black} 0.1173} & {\color{black} 0.00056542} & {\color{black} 0.0238} & {\color{black} 0.0737} & {\color{black} 0.00082948} & {\color{black} 0.0288} & {\color{black} 0.0847} & {\color{black} \textbf{0.00052629}} & {\color{black} \textbf{0.0229}} & {\color{black} \textbf{0.0691}} \\ \hline
  {\color{black} 4h} & {\color{black} 0.00113856} & {\color{black} 0.0337} & {\color{black} 0.1246} & {\color{black} \textbf{0.00071719}} & {\color{black} \textbf{0.0268}} & {\color{black} 0.0942} & {\color{black} 0.0011012} & {\color{black} 0.0332} & {\color{black} 0.1143} & {\color{black} 0.00079793} & {\color{black} 0.0282} & {\color{black} \textbf{0.0884}} \\ \hline
  {\color{black} 6h} & {\color{black} 0.00113166} & {\color{black} 0.0336} & {\color{black} 0.1087} & {\color{black} 0.00079771} & {\color{black} 0.0282} & {\color{black} 0.0886} & {\color{black} 0.00094857} & {\color{black} 0.0308} & {\color{black} 0.1010} & {\color{black} \textbf{0.0007228}} & {\color{black} \textbf{0.0269}} & {\color{black} \textbf{0.0787}} \\ \hline
  {\color{black} 8h} & {\color{black} 0.00156561} & {\color{black} 0.0396} & {\color{black} 0.1350} & {\color{black} 0.00074263} & {\color{black} 0.0273} & {\color{black} 0.0880} & {\color{black} 0.00086032} & {\color{black} 0.0298} & {\color{black} 0.0960} & {\color{black} \textbf{0.00073787}} & {\color{black} \textbf{0.0272}} & {\color{black} \textbf{0.0833}} \\ \hline
  {\color{black} 12h} & {\color{black} 0.00164631} & {\color{black} 0.0406} & {\color{black} 0.1368} & {\color{black} \textbf{0.00065455}} & {\color{black} \textbf{0.0256}} & {\color{black} 0.0819} & {\color{black} 0.00095340} & {\color{black} 0.0309} & {\color{black} 0.0954} & {\color{black} 0.00070197} & {\color{black} 0.0265} & {\color{black} \textbf{0.0814}} \\ \hline
  {\color{black} 15h} & {\color{black} 0.00170941} & {\color{black} 0.0413} & {\color{black} 0.1368} & {\color{black} 0.00086864} & {\color{black} 0.0295} & {\color{black} 0.0876} & {\color{black} 0.0010403} & {\color{black} 0.0325} & {\color{black} 0.1017} & {\color{black} \textbf{0.00073697}} & {\color{black} \textbf{0.0271}} & {\color{black} \textbf{0.0822}} \\ \hline
\end{tabular}%
}
\end{table}

    \color{black}To compare the performance of different approaches in terms of training and predicting cost, we compare the training time and prediction time as shown in Table \ref{tab:traing_time_comp}. The training time is the average time consumed for training one epoch, and the prediction time is the mean value of predicting 1000 lines of data by repeating 10 times. Based on the results, we can observe that the esDNN consumes the longest training time with about 10\% more time than Bi-LSTM and GRU, which is an acceptable cost considering the performance improvement in prediction accuracy. The longer training time can result from the more complicated model of esDNN with application of CNN. And for the prediction time, esDNN can perform slightly better than Bi-LSTM and GRU, the reason can be the Swish activation function that we use can slightly improve the prediction time \cite{Swish}. \color{black}
    
    \begin{table}[]
    	\caption{\color{black}Training time and prediction time comparison}
    	\label{tab:traing_time_comp}
    	\centering
    	
 \resizebox{0.4\textwidth}{!}{%
 	\begin{tabular}{|c|c|c|c|c|}
 		\hline
 		{\color{black} }                    & {\color{black} \textbf{RNN}} & {\color{black} \textbf{Bi-LSTM}} & {\color{black} \textbf{GRU}} & {\color{black} \textbf{esDNN}} \\ \hline
 		{\color{black} Traing time (s)}     & {\color{black} 5.11}         & {\color{black} 6.47}          & {\color{black} 6.53}         & {\color{black} 7.14}           \\ \hline
 		{\color{black} Prediction time (s)} & {\color{black} 0.048}        & {\color{black} 0.070}         & {\color{black} 0.071}        & {\color{black} 0.065}          \\ \hline
 	\end{tabular}%
 }
\end{table}

    To summarize, esDNN can achieve good accuracy based on MSE results. Compared with the MSE results evaluated in the same datasets derived from Google and Alibaba in \cite{ChenTPDS2020}, esDNN has reduced the MSE with one order of magnitude from around $7 \times 10^{-2}$ to $7\times 10^{-3}$.

	\subsection{\color{black}Applying esDNN for Machines Auto-scaling with Simulations}
	
		
\begin{figure*}[t]
	\centering
	\begin{subfigure}[b]{0.48\linewidth}
		\centering
		\includegraphics[height=4cm, width=0.9\linewidth]{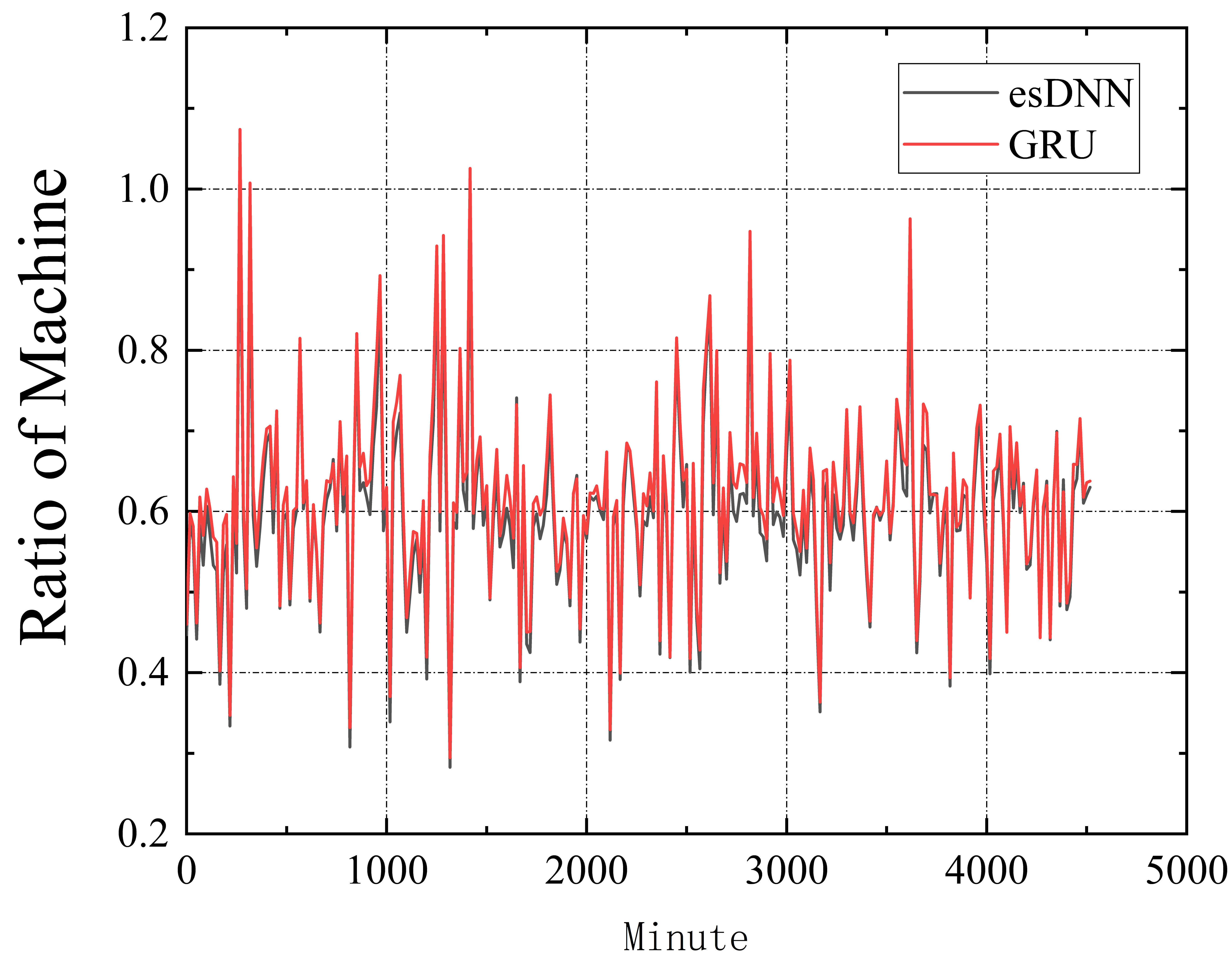}
		\caption[VarPerOptCom]{\color{black}Optimization ratio of Alibaba machines}
		\label{fig:Alibaba machine}
	\end{subfigure}
	\begin{subfigure}[b]{0.48\linewidth}
		\centering
				\includegraphics[height=4cm, width=0.9\linewidth,trim=0.1in 0.2in 0.1in 0.3in]{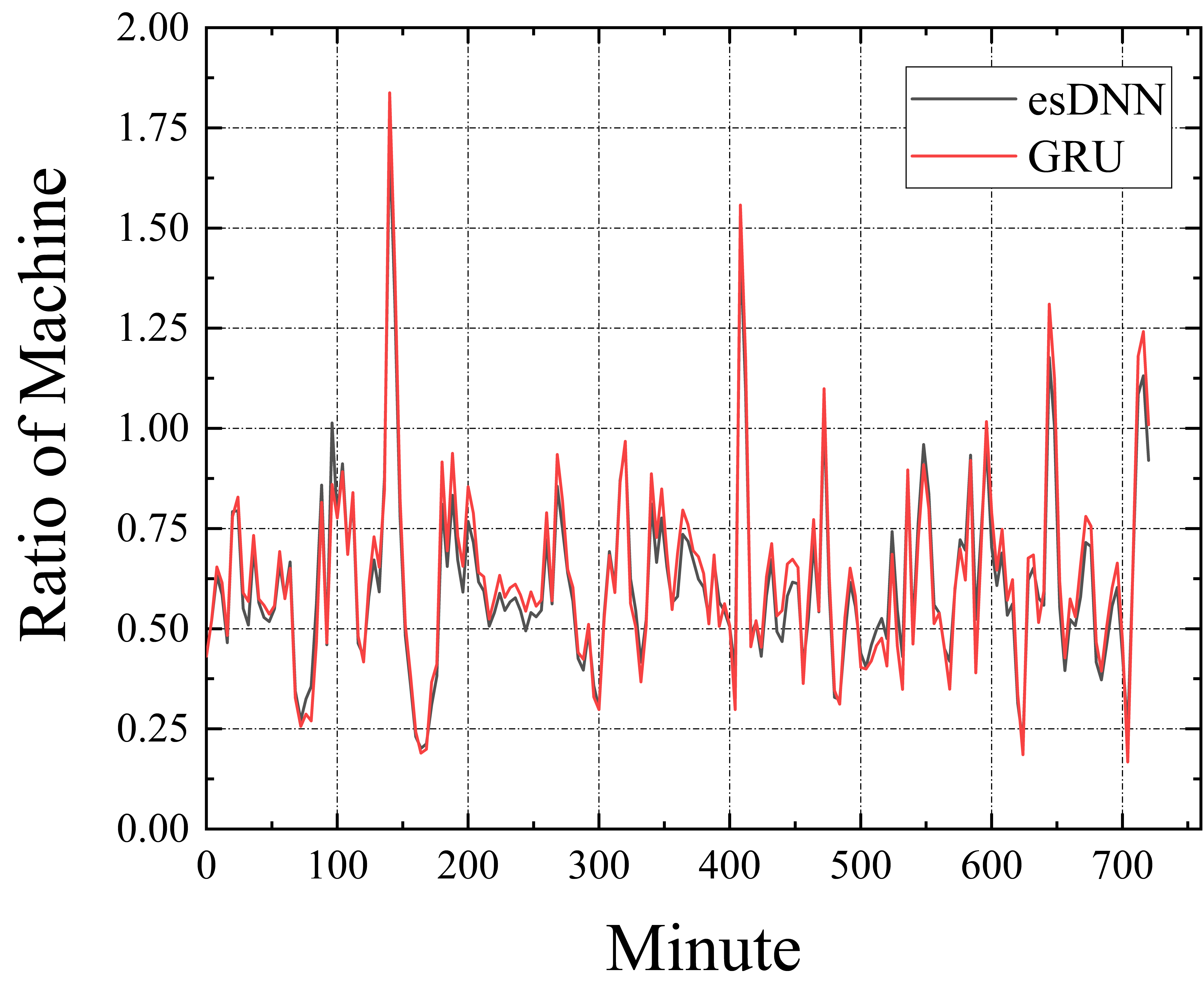}
			\caption[VarPerOptCom]{\color{black}Optimization ratio of Google machine}
			\label{fig:Google machine}
	\end{subfigure}
\caption{\color{black}Optimization ratio comparison in minutes with different traces}
\end{figure*}
	

 The auto-scaling technique can dynamically adjust the number of active machines in the system based on the system status, e.g. removing machines when the system is at a low utilization level or adding more machines when the system is overutilized. By taking advantage of auto-scaling, the system performance, e.g. energy consumption, can be optimized. However, without sufficiently accurate prediction for workloads, the popular threshold-based auto-scaling approaches, like static threshold, are undesirable for workloads with high variance.  

    \color{black}To further demonstrate the capability of the proposed approach, we integrate esDNN into the auto-scaling scenario for physical machines in Alibaba and Google cloud data centers by simulating the number of machines and resource usage. \color{black}The specifications of machines are derived from the corresponding original datasets, and the scheduling period is configured as 5 minutes. 
	
    Our objective is to improve resource utilization and reduce the number of active machines with sufficient accurate predictions. Therefore, the use of CPU utilization as an input to the auto-scaling method is highly desired. And the output is the number of active machines. As an auto-scaling baseline, we use the average number of active machines based on the previous time slots \cite{Adel2017JNCA}, which can be calculated as:
    \begin{equation}
        M(t)=\frac{\sum_{i=1}^{m} M(t-i)}{m},
        \label{no_active}
    \end{equation}
	where $M(t)$ represents the number of active machines at time interval $t$, and $m$ is the number of previous time slots used for the prediction that we set $m$ as 5 for our experiments. 
    The number of active machines calculated by Eq.~\ref{no_active} is normalized as 1.0. We normalize the number of active machines of the auto-scaling approach based on esDNN and calculate the ratio between the esDNN-based approach and baseline, and we configure the upper CPU utilization threshold as 80\%. If the ratio of machines is less than 1.0, it means that the esDNN-based approach can reduce the number of active machines. Fig.~\ref{fig:Alibaba machine} shows the prediction of the number of active machines based on the Alibaba dataset. As expected, the ratio fluctuates in the range of 0.3 to 0.6, indicating that our prediction algorithm has achieved a good effect that only 40\% to 80\% of the number of machines will be active compared with the original number of machines from the dataset, which can significantly reduce the number of active machines. 

    As for the Google dataset, we analyze the capacity distribution of about 37,678 machines, where the machines with 0.5 capacity are about 92.8\%, the machines with 0.25 capacity are about 1.4\%, and the machines with capacity 1.0 are about 5.9\%. This is different from the distribution of the homogeneous machines in the Alibaba dataset. 
    After normalizing the CPU usage in the Google dataset, we also utilize the algorithm previously applied to the Alibaba dataset, and Fig.~\ref{fig:Google machine} shows the ratio ranges from 0.25 to 1. For example, at the 400th minute, the baseline needs 18,371 machines, while our approach only uses 5,161 machines. The results of these experiments are close to those based on the Alibaba dataset. It can be concluded that the proposed approach can efficiently improve resource usage by reducing the number of active machines, and it is promising to reduce the energy consumption of cloud data centers by providing an accurate prediction method. 
	
	\section{Conclusions and Future Work}
	\label{sec:conclusion}
	Our deep learning-based approach for cloud workload prediction brings opportunities to optimize resource provisioning in the cloud computing environment. In this paper, we apply sliding window for multivariate time series to convert the high dimension data into supervised learning time series to address the high dimensionality challenge. Based on the converted data, we proposed a revised GRU-based approach to train the prediction model to achieve high prediction accuracy for high variance cloud workloads. Comprehensive experiments based on the realistic traces derived from Google and Alibaba have demonstrated that our proposed approach can achieve better performance in terms of accuracy compared with state-of-the-art approaches. To further show the effectiveness of optimizing resource provisioning, we applied our approach for auto-scaling based on realistic traces, and results illustrate that our approach can significantly optimize the resource usage of cloud data centers, thus saving operational costs. 
	
\color{black} In future,  our approach can be integrated into a container-based prototype system, e.g. Kubernetes, to optimize resource provisioning. We would like to investigate the proposed approach to be extended for Edge Computing to reduce response time using offloading techniques, and consider the location-aware and mobility-aware scenarios (e.g. predicting the loads allocating to different devices generated by mobile users). We would also like to make automatic esDNN by using Monitor, Analyze, Plan, and Execute (MAPE) model. \color{black}

	\section*{Acknowledgment}
	
This work is supported by National Key R\&D Program of China (No. 2021YFB3300200), National Natural Science Foundation of China (NO. 62102408 and 62071327), Shenzhen Basic Research Program (No. JCYJ20200109115418592),  and Youth Innovation Promotion Association CAS (2019349).

	\bibliographystyle{unsrt}
\bibliography{library}

\end{document}